*Review*

# Determination of Elastic Parameters of Lipid Membranes with Molecular Dynamics: A Review of Approaches and Theoretical Aspects

**Konstantin V. Pinigin**

A. N. Frumkin Institute of Physical Chemistry and Electrochemistry, Russian Academy of Sciences, 31/4 Leninskiy Prospekt, 119071 Moscow, Russia; piniginkv@gmail.com

**Abstract:** Lipid membranes are abundant in living organisms, where they constitute a surrounding shell for cells and their organelles. There are many circumstances in which the deformations of lipid membranes are involved in living cells: fusion and fission, membrane-mediated interaction between membrane inclusions, lipid–protein interaction, formation of pores, etc. In all of these cases, elastic parameters of lipid membranes are important for the description of membrane deformations, as these parameters determine energy barriers and characteristic times of membrane-involved phenomena. Since the development of molecular dynamics (MD), a variety of *in silico* methods have been proposed for the determination of elastic parameters of simulated lipid membranes. These MD methods allow for the consideration of details unattainable in experimental techniques and represent a distinct scientific field, which is rapidly developing. This work provides a review of these MD approaches with a focus on theoretical aspects. Two main challenges are identified: (i) the ambiguity in the transition from the continuum description of elastic theories to the discrete representation of MD simulations, and (ii) the determination of intrinsic elastic parameters of lipid mixtures, which is complicated due to the composition–curvature coupling effect.

## 1. Introduction

Lipid membranes are ubiquitous in living systems. The structural unit of these membranes is a lipid molecule that can be divided into two parts: a hydrophobic hydrocarbon tail and hydrophilic head. Lipids tend to diminish the contact of their tails with water, and therefore, lipid membranes usually exist in the form of a bilayer: two layers (monolayers) of lipid molecules, the hydrophobic tails of which are hidden inside a membrane from the contact with water. The most common example of a lipid membrane is the plasma membrane, which provides an almost impermeable shell for living cells, protecting them from the external environment. Except for plasma membranes, lipid membranes also constitute a boundary of organelles: nucleus, endoplasmic reticulum (ER), Golgi apparatus, lysosomes, mitochondria, vesicles, vacuoles, etc. In addition, lipid membranes are a part of the envelope of some viruses [1].

Experimentally, it is known that lipid membranes possess elasticity [2–6]: a tendency to preserve shape in response to applied forces and return to its original configuration when forces are removed. There is a lot of evidence that elastic properties of lipid membranes play a key role in various deformation-involved circumstances, controlling energy barriers and characteristic times: fusion and fission [7–10], membrane-mediated interaction between membrane inclusions [11–17], lipid–protein interaction [18–20], formation of pores [21,22], structure of the boundary of lipid domains [23,24], sorting of membrane proteins and peptides between different phase regions [24,25].

The elasticity of lipid membranes can be characterized by elastic energy, which specifies the energy cost required for a specific deformation. The magnitude of a particular deformation is usually called a *strain*, while the strain coefficients in the elastic energy are called *elastic moduli*. While strain is a variable quantity, elastic moduli are fixed constants and thus represent the property of a considered membrane. However, elastic moduli are not the only property, as the elastic energy can also contain spontaneous values of strain. In the following, elastic moduli and spontaneous strain are referred to as *elastic parameters* of lipid membranes. In general, elastic parameters depend on lipid composition. It is known that there is a variety of different lipids that comprise the membranes of living cells [26,27], which allows cells to adjust the lipid composition of their membranes. The determination of elastic parameters is thus important for the understanding and description of how living cells regulate the course of deformation-involved processes.

Plenty of experimental techniques exist for the determination of elastic parameters of lipid membranes [28]. Since the development of MD, various *in silico* methods also appeared for the determination of elastic parameters of simulated lipid membrane models. In MD simulations, it is possible to follow the movements of individual atoms. Therefore, although the size of MD systems is rather limited due to computational limits, MD simulations allow for considering details and controlling system parameters unattainable in experiments.

In this work, the MD approaches for the determination of elastic parameters of lipid membranes are reviewed. The purpose of this work is to describe the capabilities of various approaches, their applicability conditions, problems, advantages, and disadvantages in comparison with each other. All approaches are divided into two groups: (i) equilibrium force methods and (ii) fluctuation-based methods. The first group of methods considers average (equilibrium) values of strain and stress by applying external forces to lipid membranes. The second group, in turn, analyzes the fluctuations of different membrane characteristics from their average values. The paper is organized as follows. Section 2 provides a theoretical framework necessary for the subsequent overview of MD methods. In Sections 3 and 4, the equilibrium force methods and fluctuation-based methods are reviewed, respectively. In Section 5, the results of this review are summarized.

## 2. Elastic Theory of Lipid Membranes
### 2.1. 2D Elasticity of Lipid Membranes

On a large scale, much larger than their thickness, lipid membranes can be considered as infinitely thin fluid films and are modeled as 2D surfaces. The consideration of all quadratic geometrical invariants of a surface $S$ leads to the classical Helfrich Hamiltonian [2,29,30]:

$$W = \int_S \left[ \frac{k}{2}(K - K_0)^2 + \bar{k} K_G + \sigma \right] dS, \tag{1}$$

where $W$ is the elastic energy, $K$ is the extrinsic curvature, $K_G$ is the Gaussian curvature, $K_0$ is the spontaneous curvature, $k$ is the bending modulus, $\bar{k}$ is the Gaussian curvature modulus, and $\sigma$ is the surface tension. The first two energy contributions of Equation (1) can be written in terms of principal curvatures: each point at a surface has two normal directions along the tangent plane, at which the curvatures of normal sections (intersections of a surface with planes normal to the tangent plane) obtain maximum and minimum values, $c_1$ and $c_2$, called principal curvatures. In terms of $c_1$, $c_2$, $K = c_1 + c_2$ and $K_G = c_1 c_2$. These principal curvatures, as well as $K$ and $K_G$, represent geometrical invariants that do not depend on a surface parametrization. For a cylindrical surface, for example, one of the principal directions is along the cylinder axis, the principal curvature of which equals zero, while the other direction is along the circumference and the corresponding principal curvature equals 1/R, where R is the cylinder radius; see Figure 1. Thus, $K = 1/R$ and $K_G = 0$.

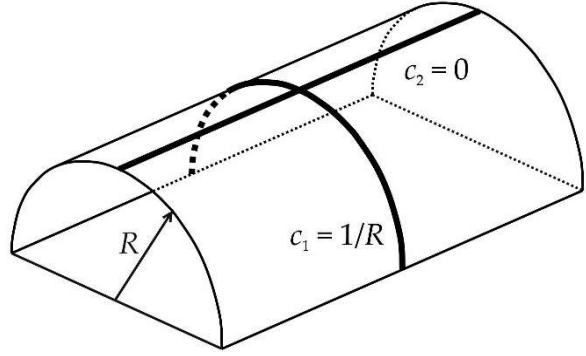

**Figure 1.** Illustration of the concept of the principal curvatures on the example of a cylindrical surface. The depicted half-cylinder of radius $R$ has two principal directions: (1) along the circumference and (2) along the cylinder axis, which are perpendicular to each other. The corresponding principal curvatures are $c_1 = 1/R$ and $c_2 = 0$.

The last term of Equation (1), the surface tension, accounts for the energy cost of pulling lipid material from a reservoir. For example, in experiments with micropipette pulling of membrane tubes, the reservoir is a lipid solution that may be employed for the formation of lipid bilayers on a grid [31]. Another example is the surface tension due to shape undulations in lipid vesicles [4], which can be considered an effective reservoir. High values of $\sigma$ can lead to a change in the area per lipid, the elastic energy of which should be additionally taken into account [3]. In MD simulations, $\sigma$ can be controlled by performing simulations in the grand canonical ensemble, wherein the virtual reservoir is characterized by some chemical potential that determines the number of lipids in a membrane; see for instance Ref. [32]. Generalizations of the Helfrich Hamiltonian exist, which additionally take into account a nematic microstructure given by a nematic in-plane director coupled to the geometry of the membrane surface [33,34]. This nematic microstructure might occur if some chiral inclusions are present in lipid membranes. For instance, a rod-like structure of cardiolipin may involve an in-plane nematic order caused by these molecules [33]. Nevertheless, in most biologically relevant cases, there is no nematic ordering in lipid membranes, and in this work, nematic-ordering parameters will not be considered.

*2.2. 3D Elasticity of Lipid Membranes*

2.2.1. Basic Assumptions

In the Helfrich Hamiltonian, the internal structure of lipid membranes is not taken into account. The internal structure, however, is important for the consideration of membrane deformations at a nanoscale, i.e., at a scale compatible with membrane thickness. Such nanoscopic deformations occur in a variety of cases: in the vicinity of membrane inclusions (membrane proteins, antimicrobial peptides, regulatory lipids) [11–17], formation of pores [21,22], stalk formation during fusion and fission [7,8], at the boundary of liquid-ordered domains (rafts) [23,24].

To take into account the internal structure of a lipid bilayer membrane, each monolayer of a lipid bilayer can be considered a three-dimensional elastic body. As a reference configuration, from which the energy is calculated, a planar state is usually chosen. This reference configuration allows for the incorporation of the following features of lipid monolayers: (i) transverse isotropy, (ii) lateral fluidity, (iii) incompressibility. The first property follows from a symmetry argument: the properties of a planar lipid monolayer are the same in every direction along the plane of a monolayer surface. The second feature is implied by the lateral fluidity of lipid molecules that can freely diffuse along the plane of a monolayer. The third feature is supported by experiments [35–38] showing a rather high bulk modulus of lipid membranes. The bulk modulus is defined as:

$$k_V = -V \frac{dP}{dV}, \qquad (2)$$

where $V$ is the total volume and $P$ is the external pressure. The bulk modulus of water for instance is 2.2 GPa at 25 °C and 0 bar [39]. Experiments show that the bulk modulus of lipid membranes is close to that of water [35–38]. This implies that to change the volume of a lipid membrane by a small amount, say 1%, it is necessary to apply a rather high external pressure of 22 MPa = 220 bar. Therefore, lipid membranes are usually assumed to be incompressible.

With these three properties taken into account, the classical theory of elasticity leads to the following quadratic energy density (energy per unit volume) of a lipid monolayer [14,40]:

$$w_{3D}^{mono} = \sigma_0(z)\varepsilon(x,y,z) + \frac{1}{2}E(z)\varepsilon(x,y,z)^2 + \frac{1}{2}\lambda_T(z)\mathbf{T}(x,y,z)^2. \quad (3)$$

In Equation (3), $x$, $y$, $z$ are the coordinates of a Cartesian coordinate system, which is chosen in such a way that the $z$-axis points from the hydrophilic to the hydrophobic part along the axis of symmetry of a monolayer. Equation (3) is written a little differently than in Refs. [14,40]. Firstly, the dependence on $x$ and $y$ was omitted in Refs. [14,40] for brevity. Secondly, although in Equation (3), $\mathbf{T}$ has the same prefactor as in Ref. [14], in Equation (3) of Ref. [40], the prefactor is $2\lambda_T(z)$, which is likely a typo, as in Equation (14) of Ref. [40], the prefactor is $\frac{1}{2}\lambda_T(z)$, the same as in Equation (1) of Ref. [41] of the same authors. Elastic strains in Equation (3) are given by two functions: $\varepsilon(x,y,z)$ and $\mathbf{T}(x,y,z)$. The first one is a scalar function that represents the local lateral area expansion: any infinitesimal lateral area $dA$ changes by $\varepsilon dA$ upon a deformation. The second one is a vector field, called local tilt, which quantifies the transverse shear deformation mode, i.e., the deviation of an average orientation of lipid molecules from the local monolayer normal. The average orientation of lipid molecules is called a lipid director and is given as a field of unit vectors, $\mathbf{n}(x,y,z)$, pointing from the hydrophilic to the hydrophobic part of a monolayer. Typically, $\mathbf{n}(x,y,z)$ is assumed to be a constant function of $z$. The local monolayer normal, $\mathbf{N}(x,y,z)$, is defined as a unit normal to a surface given by $\mathbf{X}(x,y,z)$ with fixed $z$, where $\mathbf{X}(x,y,z)$ is a deformation field of a lipid monolayer from the reference state to some deformed state. The tilt degree of freedom can be expressed through $\mathbf{n}(x,y,z)$ and $\mathbf{N}(x,y,z)$ as $\mathbf{T}(x,y,z) \equiv \dfrac{\mathbf{n}(x,y,z)}{\mathbf{N}(x,y,z) \cdot \mathbf{n}(x,y,z)} - \mathbf{N}(x,y,z)$ [14,40], where the dot sign is the scalar product. Thus, the tilt field, $\mathbf{T}(x,y,z)$, represents the degree of deviation of the director field, $\mathbf{n}(x,y,z)$, from the local normal, $\mathbf{N}(x,y,z)$ (see Figure 2).

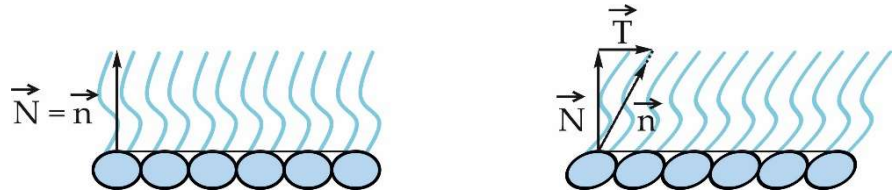

**Figure 2.** Definition of the vector fields **n**, **N** and **T**. On the left-hand side, the lipid monolayer is in the reference state, and the directions of the normal **N** to the monolayer and director **n**, which characterizes the mean orientation of lipid tails, coincide. On the right-hand side, the monolayer is tilted, and the tilt vector **T** represents the degree of deviation of **n** from **N**.

Note that Equation (3) lacks the lateral shear deformation mode, reflecting the lateral fluidity of lipid monolayers. In addition, there is only the local lateral strain given by $\varepsilon(x,y,z)$ in Equation (3), while the local longitudinal strain, $\varepsilon_z \equiv \mathbf{n} \cdot \nabla_z \mathbf{X} - 1$, is absent.

The latter follows from the incompressibility assumption. Here, it is necessary to distinguish between two types of incompressibility: local and global. Local incompressibility implies that the volume of any infinitesimal monolayer patch remains constant during deformations. From experiments, however, it is only known that lipid monolayers are incompressible in the global sense: the total volume of lipid monolayers is constant, whereas local volumes may change [35–38]. Actually, MD simulations show that local incompressibility does not hold in lipid monolayers [42,43]. Despite this, in theory, it is usually assumed for simplicity that lipid monolayers are locally incompressible, which implies that $(1+\varepsilon)(1+\varepsilon_z) = 1$ [44], i.e., $\varepsilon$ and $\varepsilon_z$ are not independent, and permits the incorporation of only $\varepsilon$ to the elastic energy. In general, however, $\varepsilon$ and $\varepsilon_z$ can be independent quantities. Nevertheless, even if global incompressibility is assumed, it was shown that $\varepsilon$ and $\varepsilon_z$ are still not independent in planar configurations of a lipid monolayer [43]. Elastic moduli in Equation (3) are given by two functions $E(z)$ and $\lambda_T(z)$. In Equation (3), $E(z)$ is a coefficient of $\dfrac{\varepsilon(x,y,z)^2}{2}$ and thus represents the local stretching modulus of a lipid monolayer. $\lambda_T(z)$ in turn multiplies $\dfrac{\mathbf{T}(x,y,z)^2}{2}$ and has the meaning of the local tilt modulus. Both $E(z)$ and $\lambda_T(z)$ do not depend on $x$ and $y$, for a lipid monolayer is transversely symmetric in the reference configuration. In contrast to the tilt degree of freedom, the local lateral stretching also has a linear term in Equation (3), given by $\sigma_0(z)\varepsilon(x,y,z)$. $\sigma_0(z)$ thus represents the local lateral pressure (or stress, i.e., minus the pressure) profile of a lipid monolayer in the reference configuration. Although $\sigma_0(z)$ can be measured experimentally [45,46], the experimental precision is not as high as can be provided by MD simulations [47,48]. In general, the stress profile has the following characteristics: a high negative pressure at the interface between the hydrophilic and hydrophobic parts of a monolayer, i.e., in the proximity of the glycerol region of lipid molecules, and mainly a repulsive positive pressure in the head-group region and hydrophobic tails [47,48].

2.2.2. Dimensional Reduction

Both the Helfrich Hamiltonian and Equation (3) are phenomenological relations between the elastic energy and strain since they are mainly based on symmetry arguments and do not rely on first principles. However, the level of phenomenology is different: in Equation (3), the third dimension is additionally considered as opposed to the Helfrich Hamiltonian, wherein lipid membranes are considered as infinitely thin fluid films. Therefore, Equation (3) is more general than Equation (1), and Equation (1) can be obtained from Equation (3) by performing a dimensional reduction from 3D to 2D. To do this, one can integrate out all energy contributions along the thickness of a monolayer. In this way, within the assumption that $\mathbf{n}(x,y,z)$ and $\mathbf{T}(x,y,z)$ does not depend on $z$, Hamm and Kozlov (HK in the following) derived the following quadratic energy functional [44]:

$$w_{2D}^{\text{mono}} = \frac{1}{2} k_{A,m} \alpha^2 + \frac{1}{2} k_m (\tilde{K} + K_{0,m})^2 + \bar{k}_m \tilde{K}_G + \frac{1}{2} k_t \mathbf{T}^2, \tag{4}$$

where:

$$k_{A,m} = \int_{m_0} E(z) dz, \quad k_m = \int_{m_0} E(z)(z - z_0)^2 dz,$$
$$-k_m K_{0,m} = \int_{m_0} \sigma_0(z)(z - z_0) dz, \quad \bar{k}_m = \int_{m_0} \sigma_0(z)(z - z_0)^2 dz, \tag{5}$$
$$k_t = \int_{m_0} \lambda_T(z) dz.$$

In Equation (5), $\int_{m_0} dz$ is the integration along the thickness from the hydrophilic to the hydrophobic part of a lipid monolayer in the reference configuration, $k_{A,m}$ is the stretching modulus, $k_m$ is the bending modulus, $K_{0,m}$ is the spontaneous curvature, $z_0$ is the position of the neutral surface, $\bar{k}_m$ is the Gaussian curvature modulus, and $k_t$ is the tilt modulus. In Equation (4), stretching $\alpha$ and tilt $\mathbf{T}$ correspond to $\varepsilon(x,y,z)$ and $\mathbf{T}(x,y,z)$ evaluated at the neutral surface, $z = z_0$. $\tilde{K}$ and $\tilde{K}_G$, which are defined on the neutral surface, are called the effective extrinsic curvature and effective Gaussian curvature, respectively. They are constructed from the effective curvature tensor: $\tilde{K}_{ij} = K_{ij} + \nabla_i T_j$, where $\nabla_i$ is the covariant derivative along the $i$-th coordinate ($x$ and $y$ refer to the first and second coordinate, respectively), $K_{ij} = -\mathbf{N} \cdot \nabla_i \mathbf{e}_j$ (where $\mathbf{e}_j = \nabla_j \mathbf{X}$ are local basis vectors) is the second fundamental form of the neutral surface, $T_j = \mathbf{T} \cdot \mathbf{e}_j$ are the covariant components of the tilt vector. In these notations, $\tilde{K} = g^{ij}\tilde{K}_{ij} = K + \boldsymbol{\nabla} \cdot \mathbf{T}$ and $\tilde{K}_G = \dfrac{\tilde{K}_{11}\tilde{K}_{22} - \tilde{K}_{12}\tilde{K}_{21}}{g}$, where $g^{ij}$ is the inverse of the metric tensor $g_{ij} = \mathbf{e}_i \cdot \mathbf{e}_j$ and $g = \det(g_{ij}) = g_{11}g_{22} - g_{12}^2$. Although the expressions for $\tilde{K}$ and $\tilde{K}_G$ in terms of $\tilde{K}_{ij}$ are rather cumbersome, the actual meaning of $\tilde{K}$ and $\tilde{K}_G$ is more transparent when written in terms of the lipid director $\mathbf{n}$. At any point of the neutral surface, a local Cartesian coordinate system can be constructed with the origin at this point and the $z$-axis pointed along $\mathbf{N}$ in the same direction. Then, $\tilde{K} \approx \dfrac{\partial n_x}{\partial x} + \dfrac{\partial n_y}{\partial y}$, i.e., approximately equals the divergence, $\boldsymbol{\nabla} \cdot \mathbf{n}$, of the lipid director; $\tilde{K}_G \approx \dfrac{\partial n_x}{\partial x}\dfrac{\partial n_y}{\partial y} - \dfrac{\partial n_x}{\partial y}\dfrac{\partial n_y}{\partial x} = \det\left(\dfrac{\partial n_i}{\partial x_j}\right)$, where $(x_1, x_2) = (x, y)$. The position of the neutral surface, $z_0$, is defined by the equation $\int_{m_0} E(z)(z - z_0) dz = 0$, which implies that the stretching and bending modes are decoupled at this surface. An equivalent definition of the neutral surface is that it is a surface with respect to which the bending modulus is minimal. The bending modulus relative to an arbitrary plane $\tilde{z} = \text{const}$ can be written as $k_m(\tilde{z}) = \int_{m_0} E(z)(z - \tilde{z})^2 dz$ [14]. The extremum of $k_m(\tilde{z})$ satisfies $\dfrac{\partial k_m(\tilde{z})}{\partial \tilde{z}} = -2\int_{m_0} E(z)(z - \tilde{z}) dz = 0$, which coincides with the definition of the neutral surface. That this extremum corresponds to the minimum follows from the equality $\dfrac{\partial^2 k_m(\tilde{z})}{\partial \tilde{z}^2} = 2\int_{m_0} E(z) dz = 2k_A > 0$. Thus, at $\tilde{z} = z_0$ the bending modulus is minimum. This implies that when an arbitrary torque is applied, a lipid monolayer would bend relative to the neutral surface to minimize the elastic energy. Therefore, the bending modulus relative to the neutral surface is of most interest for the determination.

When $\mathbf{T} = 0$, the HK Hamiltonian, Equation (4), transforms to the Helfrich Hamiltonian amended by the elastic energy of stretching. At the macroscopic scale, the stretching deformation mode, being a rather tough one in comparison with bending, is usually omitted [2]. Thus, the HK Hamiltonian represents a generalization to the Helfrich Hamiltonian for lipid monolayers, additionally taking into account the tilt deformation mode, i.e., the deviation of the average orientation of lipid molecules from the local normal to the monolayer surface. Although the signs of the spontaneous curvatures in Equations (3) and (4) are different, this is a matter of definition. Equation (4) follows the definition usually used in experiments [49,50]: a positive spontaneous curvature implies that lipid tails tend to repel more strongly than lipid heads.

The HK Hamiltonian, written in the form of Equation (4), relies on the local fluidity assumption, $\lambda_{LS}(z) = 0$, where $\lambda_{LS}(z)$ is the lateral shear modulus. However, a more general assumption can be assumed, the global fluidity assumption: $\int_{m_0} \lambda_{LS}(z) dz = 0$. Actually, global fluidity is enough to take into account the free translational motion of lipids in the lateral direction, which is usually meant by the lateral fluidity of lipid monolayers. Along with local fluidity, HK also considered global fluidity and derived that, in the case of the global fluidity assumption, Equation (4) should be supplemented by the twist term [44]: $\frac{1}{2} k_{tw} (\nabla \times \mathbf{T})^2$, where $k_{tw} = \int_{m_0} \lambda_{LS}(z)(z-z_0)^2 dz$ and $\nabla \times \mathbf{T} = \frac{\nabla_1 T_2 - \nabla_2 T_1}{\sqrt{g}}$. In addition, it turns out that $\lambda_{LS}(z)$ modifies the bending and Gaussian curvature moduli: $k_m \to k_m + k_{tw}$, $\bar{k}_m \to \bar{k}_m - 4k_{tw}$ [44]. The contribution of $\lambda_{LS}(z)$ to $\bar{k}_m$ was also obtained in Ref. [51]. In a subsequent elaboration of the HK theory, it was argued [14] that stability considerations require $\lambda_{LS}(z) \geq 0$, and thus, the global and local fluidity assumptions are equivalent: from $\lambda_{LS}(z) \geq 0$ and $\int_{m_0} \lambda_{LS}(z) dz = 0$ it follows that $\lambda_{LS}(z) = 0$. Moreover, it was shown that within the local fluidity assumption, the HK Hamiltonian, Equation (4), is unstable: it is possible to construct a deformation that takes the elastic energy to minus infinity, leading to unphysical results [14]. This instability occurs due to the effective Gaussian curvature term, $\tilde{K}_G$, in Equation (4), and it was suggested that this term should be omitted from the elastic energy [14]. Another argument to omit $\tilde{K}_G$ stems from the fact that the Gaussian curvature modulus, $\bar{k}_m$ equals $\int_{m_0} \sigma_0(z)(z-z_0)^2 dz$ (see Equation (5)) and thus contains only the pre-stress $\sigma_0(z)$. In general, however, in any quadratic theory, pre-stress contributions to the elastic energy should contain only linear terms.

In the HK theory, the local tilt field, $\mathbf{T}(x,y,z)$, was assumed to be a constant function of z. In subsequent developments of the HK theory [14,40,41], a more general case was considered: the dependence of $\mathbf{T}(x,y,z)$ on z was taken into account, which led to the following quadratic energy functional [14]:

$$e_{2D}^{mono} = \frac{1}{2} k_{A,m} \alpha^2 + \frac{1}{2} k_m (\tilde{K} - K_{0,m})^2 + \frac{1}{2} k_t \mathbf{T}^2$$
$$+ k_c \mathbf{T} \cdot \nabla \tilde{K} + \frac{k_{gr}}{2} (\nabla \tilde{K})^2 + B \mathbf{T} \cdot \nabla \alpha - k_c (\nabla \alpha)^2 + C \nabla \alpha \cdot (\nabla \tilde{K}),$$
(6)

where:

$$k_c = -\frac{1}{2} \int_{m_0} \lambda_T(z)(z-z_0)^2 dz, \quad k_{gr} = \frac{1}{4} \int_{m_0} \lambda_T(z)(z-z_0)^4 dz,$$
$$B = -\int_{m_0} \lambda_T(z)(z-z_0) dz, \quad C = \frac{1}{2} \int dz \, \lambda_T(z)(z-z_0)^3 dz.$$
(7)

The HK Hamiltonian is thus supplemented by additional energy contributions that characterize the coupling between the tilt field **T**, stretching $\alpha$ and effective curvature $\tilde{K}$. The tilt–curvature coupling term, for example, is given by $k_c \mathbf{T} \cdot \nabla \tilde{K}$, where $k_c$ is the tilt–curvature coupling modulus. The $\mathbf{T}-\alpha$ and $\alpha - \tilde{K}$ couplings are given by $B\mathbf{T} \cdot \nabla \alpha$ and $C \nabla \alpha \cdot (\nabla \tilde{K})$, respectively. In addition, Equation (6) contains the contributions from the effective curvature gradient, $\frac{k_{gr}}{2} (\nabla \tilde{K})^2$, and stretching gradient, $-k_c (\nabla \alpha)^2$. This stems from the fact that even if at the neutral surface $\mathbf{T}(z=z_0)=0$, $\mathbf{T}(z \neq z_0)$, in general, is not zero; see Figure 1 of Ref. [14]. Although in Refs. [40,41] the $(\nabla \tilde{K})^2$ term was not included in the elastic energy, in Ref. [14], it was shown that this term is necessary for the stability of a membrane, as $(\nabla \tilde{K})^2$ enters the expression for $\mathbf{T}(z)^2$ in Equation (3) [14]:

$$\mathbf{T}(z)^2 = \left[\mathbf{T}(z=z_0) - (z-z_0)\nabla\alpha - \frac{1}{2}(z-z_0)^2 \nabla\tilde{K}\right]^2. \tag{8}$$

The left-hand side of Equation (8) is always ≥ 0, while the right-hand side can become negative if $(\nabla \tilde{K})^2$ is neglected after expanding the brackets, causing membrane instability [14]. Note that when $\mathbf{T} = \alpha = 0$, Equation (6) does not transform the Helfrich Hamiltonian, since $\frac{k_{gr}}{2}(\nabla K)^2$ remains. However, the $(\nabla K)^2$ term is not necessary for the stability of the Helfrich Hamiltonian, as in the Helfrich Hamiltonian, the transverse shear deformation mode is not taken into account. In addition, the Helfrich Hamiltonian has a definite stability condition, $-2k \leq \bar{k} \leq 0$ [14,52]. In ref. [14], it was shown that the values of the moduli given in Equation (7) can influence the character of membrane-mediated interactions between peripheral membrane inclusions and thus may have biological implications.

## 3. Equilibrium Force Methods

Applying forces of different magnitudes to an elastic material and measuring corresponding strains, it is possible to establish a stress–strain relation: an important characteristic that allows for the determination of elastic parameters of this material. This section is about the MD approaches that consider the stress–strain relation of lipid membranes by measuring average equilibrium values of applied forces and corresponding strains. The other group of methods, which are based on the analysis of fluctuations of forces and strains from their average values, is considered in Section 4.

Before proceeding further, there are multiple scales in terms of resolution that can be performed in MD simulations. The most detailed representation is provided by all-atomic MD simulations, in which each atom of a considered system is modeled by one bead. However, all-atomic systems usually require large computational resources. Therefore, many models have been developed wherein one bead represents multiple real atoms at once, which significantly reduces computational costs. These models, called coarse-grained models, can have various granularity levels. For example, in a coarse-grained model called Martini [53], frequently employed for MD simulations of biomolecular systems, one water bead represents four water molecules. Within the Martini model, dipalmitoylphosphatidylcholine (DPPC), which consists of 130 atoms, is represented by 12 beads. In one of the other models of lipid membranes, due to Cooke et al. [54,55], lipids are represented by just three beads. In Ref. [56], within the dissipative particle dynamics [57], lipids were represented by two beads: one hydrophobic and one hydrophilic. In Refs. [58,59], a further coarse-graining was employed with lipid membranes being represented by a one-dimensional meshless film of interacting particles. It is also a common practice to switch different degrees of granularity during one simulation, mapping and then remapping all-atomic models to coarse-grained ones to speed up some long-lasting processes, which is called multiscale modeling [60–63]. In all of these models, the force field parameters are tuned in such a way as to achieve similarity between a simulating model and the physical properties of the corresponding real system. Although all-atomic models provide the most comprehensive description of real systems, the parameters of coarse-grained models are usually finely tuned to achieve a close coincidence of emergent physical properties with known experimental data [53,64]. The MD approaches for the determination of elastic parameters of lipid membranes, considered in the following, can be applied to almost all MD models, regardless of the granularity level, and the applicability of a particular method to different types of MD models will be mentioned if necessary.

*3.1. Planar Lipid Bilayers*

Although the planar configuration of lipid bilayers, see Figure 3A, is rather simple, quite a few elastic parameters can be determined in this configuration. This section is divided into two parts: macroscopic stress methods and microscopic stress methods.

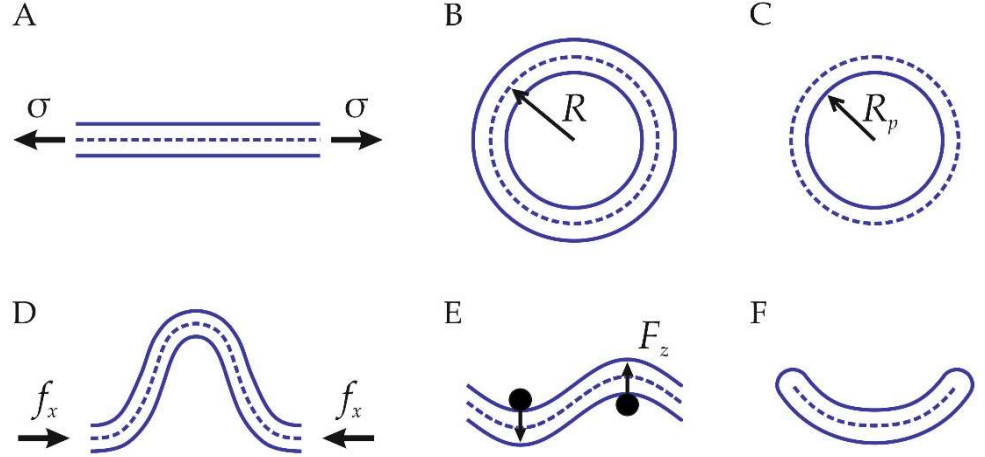

**Figure 3.** Major configurations of lipid membranes, considered in Section 3. Solid and dashed curves represent the shapes of lipid monolayers and midsurfaces, respectively. (**A**) Planar configuration of a lipid bilayer, subjected to the lateral tension $\sigma$. (**B**) Cross section of a bilayer nanotube of radius $R$. (**C**) Cross section of a cylindrical micelle, an individual cell of the H$_{II}$ phase; $R_p$ is the distance from the pivotal surface to the cylinder center. The dashed circle represents the boundary of the hydrophobic region of lipid tails. (**D**) Membrane buckle due to a large lateral compression with force $f_x$. (**E**) Sinusoidal configuration, generated by two parallel cylinders (black circles), each of which pushes the membrane with the force of magnitude $F_z$. (**F**) Unbending configuration of a half-cylinder.

3.1.1. Macroscopic Stress Methods

Due to the symmetry of the planar configuration, the local stress tensor of a planar bilayer in the Cartesian coordinate system, the *z*-axis of which is normal to the bilayer, can be written as:

$$\Sigma(x,y,z) = \begin{pmatrix} S_L(z) & 0 & 0 \\ 0 & S_L(z) & 0 \\ 0 & 0 & S_z(z) \end{pmatrix}, \qquad (9)$$

where $S_L$ and $S_z$ are the lateral and longitudinal stresses, respectively, which do not depend on *x*, *y* due to the transverse isotropy of the planar configuration. The sign convention employed for $\Sigma$ is the following: for any volume inside a lipid bilayer and for any unit vector $\mathbf{v}$ normal to the surface of this volume, $\Sigma \mathbf{v}$ equals the force per unit area, which acts on this volume. Stability requirements demand that the divergence of $\Sigma$ equals zero: $\frac{\partial S_L}{\partial x} = \frac{\partial S_L}{\partial y} = \frac{\partial S_z}{\partial z} = 0$, implying that $S_z$ does not depend on *z*. MD simulations allow for setting up two pressure values: the longitudinal pressure, $P_z$, and lateral pressure, $P_L$, which are equal to 1 bar in normal conditions. With given $P_L$ and $P_z$, $S_z(z) = -P_z = \text{const}$ and $P_L = -\frac{1}{L_z} \int_{\text{box}} S_L(z) dz$, where $L_z$ is the side length of the simulation box along the z-axis and $\int_{\text{box}} dz$ is the integration over the simulation box along the z-axis.

Perhaps, the simplest parameter to determine in the planar configuration of lipid membranes is the stretching modulus, $k_{A,m}$. The elastic energy density per unit area for stretching $\alpha$ of a lipid bilayer can be written as:

$$w_A = \frac{1}{2} k_A \alpha^2, \tag{10}$$

where $k_A$ is the bilayer stretching modulus. In terms of the monolayer stretching modulus, $k_A = 2k_{A,m}$, i.e., twice the contribution from individual monolayers; see Equation (4). The derivative of Equation (10) with respect to $\alpha$ equals:

$$\sigma(\alpha) \equiv k_A \alpha, \tag{11}$$

which is called lateral tension. Note that $\sigma(\alpha)$ is different from that of the Helfrich Hamiltonian, Equation (1), which corresponds to the tension due to some reservoir and corresponds to the energy cost required to change the membrane area with a fixed area per lipid. $\sigma(\alpha)$ of Equation (11), in turn, accounts for the change in the area per lipid. Since $\sigma(\alpha) d\alpha$ represents an infinitesimal work, $\sigma(\alpha)$ can be determined, within the incompressibility assumption, from $S_L(z)$ and $S_z$ as [65]:

$$\sigma(\alpha) = \int_{\text{box}} (S_L(z) - S_z) dz = L_z(\alpha)(P_z - P_L(\alpha)), \tag{12}$$

where $P_L(\alpha)$ and $L_z(\alpha)$ are the lateral pressure and side-length of the simulation box at stretching $\alpha$, respectively. From Equation (11), it follows that $k_A$ can be determined in MD simulations from the slope of $\sigma(\alpha)$ as a function of $\alpha$ [66–68]:

$$k_A = \frac{d}{d\alpha} \sigma(\alpha). \tag{13}$$

This approach, however, is complicated by in-plane undulations: due to thermal motion, lipid bilayers constantly undergo bending deformations, and at any given time, the membrane is not flat. The latter implies that when a planar bilayer is stretched, the projected area may increase by reducing the magnitude of undulations with the area per lipid remaining constant. This effect is observed both in experiments [3,4,69] and MD simulations [70–72]. Formally, these undulations can be taken into account by introducing an additional deformation mode. The analysis of undulations of flat bilayer membranes leads to [3,30]:

$$\alpha' = \frac{k_B T}{8\pi k} \ln\left(1 + \frac{\sigma A}{\pi^2 k}\right), \tag{14}$$

where $A$ is the initial projected area of a bilayer, and $\alpha'$ is stretching of the projected area due to undulations. The physical idea behind Equation (14) is by employing the equipartition theorem for the Fourier amplitudes of shape undulations to calculate the average excess surface area over the projected area. The amplitude of these undulations is restricted by two factors: the surface tension and elastic energy of bending, which leads to the presence of both $\sigma$ and $k$ in Equation (14). It can be assumed that (by analogy with springs in series) the total strain, $\tilde{\alpha}$, is a sum of $\alpha$ and $\alpha'$ [30]:

$$\tilde{\alpha} = \alpha + \alpha' = \alpha' = \frac{k_B T}{8\pi k} \ln\left(1 + \frac{\sigma A}{\pi^2 k}\right) + \frac{\sigma}{k_A}. \tag{15}$$

Since there is not only the stretching modulus $k_A$ but also the bending modulus $k$ in Equation (15), it follows that by fitting Equation (15) to simulation data, one can obtain both $k_A$ and $k$ as was obtained in Refs. [71,72]. The same approach is employed in experiments [3,4,69], where the undulations effect is more prominent. Actually, the smaller the system, the smaller the effect of undulations on stretching. $\alpha' \approx \frac{k_B T \sigma A}{8\pi^3 k^2}$ as $\sigma \to 0$, leading to the effective spring constant of undulations equal to $k_{Au} = \frac{8\pi^3 k^2}{A k_B T}$. Therefore, the crossover membrane area $A_c$, which can be estimated from the equality $k_{Au} = k_A$, is $A_c = \frac{8\pi^3 k^2}{k_A k_B T}$. For the typical values of $k \approx 20 k_B T$ and $k_A \approx 60 k_B T / \text{nm}^2$ [69], $\sqrt{A_c} \approx 40$ nm. This implies that for box side lengths larger than 40 nm, stretching at small tension is largely dominated by the reduction of undulations. For example, for lipid vesicles of ~20 μm in diameter, the crossover tension was found to be ~0.12 $k_B T / \text{nm}^2$ [3]. For smaller systems, the area per lipid instead is mostly involved; see Figure 2 of Ref. [71]. The theoretical prediction for stretching due to undulations, Equation (14), depends on a power spectrum of shape fluctuations [30]. The latter however often deviates from theoretical predictions at short wavelengths (see Section 4.3), which complicates a direct derivation of the expression for the undulations-involved stretching.

Apart from the stretching modulus, another important parameter, which can be determined from the macroscopic stress measurements, is the bulk modulus $k_V$, given by Equation (2). A direct application of Equation (2) would be the measurement of the total volume under the applied external pressure:

$$k_V = \frac{\ln\left(\frac{V_1}{V_2}\right)}{P_2 - P_1}. \tag{16}$$

Equation (16) was employed in Ref. [73], wherein $P_1$ and $P_2$ were chosen to be 0 and 0.1 kbar, respectively. The external pressure was applied to lipid bilayers along the z-axis with the lateral sides of the simulation box being fixed, and $V$ was calculated as a product of a bilayer area and thickness, which was defined as the distance between the phosphate atoms. The obtained values of $k_V$ were 0.35 and 0.21 GPa for atomistic and coarse-grained models, respectively. These are quite large values. From Equation (2), it follows, for example, that to change the bilayer thickness by 1%, it is necessary to apply pressures as high as 35 and 21 bar, respectively. Therefore, most often, lipid membranes are assumed to be incompressible.

An indirect way to determine $k_V$ was suggested in Ref. [74]. A lipid bilayer can be simulated at different values of the cutoff radius, $r_c$, of the Lennard–Jones interactions, which results in different total volumes of a simulated system. It can be shown that the volume correction $\Delta V$ due to a finite value of $r_c$ is:

$$\Delta V = -\frac{1}{k_V} \frac{16\pi N^2}{3V} \frac{\epsilon \sigma^6}{r_c^3}, \tag{17}$$

where $\epsilon$ and $\sigma$ are the Lennard–Jones parameters, $N$ is the number of particles, and $V$ is the total volume. The value of $k_V$, measured in this way, was $\approx 2.2$ GPa. This value approximately equals that of water [39] and is close to experimental results [35–38]. As was pointed out in Ref. [74], the application of Equation (17) has one caveat: interactions are not limited to only that of Lennard–Jones but also include electrostatic ones. However, the influence of electrostatics interactions on Equation (17) can be assumed to be small, as the number of Lennard–Jones interactions is much larger than that of electrostatic ones [74].

### 3.1.2. Local Stress Methods

These methods require information about the local characteristics of the lateral stress profiles, not only their mean values, and are based on Equation (5) of the theory section. For example, from the first moment of the stress profile, it is possible to determine the product of the bending modulus and spontaneous curvature: $-k_m K_{0,m} = \int_{m_0} \sigma_0(z)(z-z_0)dz$. For a symmetric lipid bilayer in a tensionless state, $\int_{m_0} \sigma_0(z)dz$ equals zero, and we can write:

$$-k_m K_{0,m} = \int_{m_0} \sigma_0(z) z \, dz . \tag{18}$$

The right-hand side of Equation (18) equals the force moment per unit length along the monolayer surface due to the pre-stress $\sigma_0(z)$, which implies that if this force moment is nonzero, the monolayer tends to take a shape of a certain nonzero extrinsic curvature given by the left-hand side of Equation (18). Although Equation (18) provides only the product of $k_m$ and $K_{0,m}$, Equation (18) still contains a lot of useful information. In particular, as $k_m$ is always positive, from Equation (18), it is possible to determine the sign of the monolayer spontaneous curvature. In addition, as was shown in Ref. [75], using Equation (18), it is possible to infer the additivity or nonadditivity of the spontaneous curvature as a function of composition. In particular, from the nonadditivity of the first moment of the stress profile, it was inferred in Ref. [75] that the spontaneous curvature of two lipids, sphingomyelin and cholesterol, abundant in biological membranes, is nonadditive. This implies that the additive model, which is usually assumed in experiments [49,50], may be incorrect for the description of the regulation of the spontaneous curvature of biological membranes by living cells.

Equation (18) can also be applied to lipid bilayers, in which case Equation (18) transforms to:

$$-k K_{0,b} = \int_{b_0} \sigma_0(z) z \, dz , \tag{19}$$

where $\int_{b_0} dz$ is the integration over the bilayer thickness, $k$ and $K_{0,b}$ are the bilayer bending modulus and spontaneous curvature, respectively. For symmetric bilayers, the right-hand side of Equation (19) equals zero, implying that $K_{0,b}$ is also zero. However, for asymmetric bilayers, $K_{0,b}$ can be nonzero. The latter may occur due to a different leaflet composition or different concentrations of adsorbed molecules on opposite sides of a bilayer. In Ref. [76], it was shown that the right-hand side of Equation (19) can be expressed in terms of a difference in the leaflet lateral tensions $\Delta\sigma$:

$$-k K_{0,b} = l_{sy} \Delta\sigma , \tag{20}$$

where $l_{sy}$ is the effective leaflet thickness of a symmetric bilayer. In addition, it was shown that at low coverage of adsorbed molecules $\Delta\sigma \approx k_B T \Delta\Gamma$, where $\Gamma$ is the concentration of adsorbed particles, and therefore, the following holds:

$$-k K_{0,b} = k_B T l_{sy} \Delta\Gamma . \tag{21}$$

A precise determination of $l_{sy}$ requires the fitting of $l_{sy}\Delta\sigma$ values plotted against the right-hand side of Equation (19) [76]. However, $l_{sy}$ can also be estimated in a reasonable way, which allows for the determination of $K_{0,b}$ from Equation (21) (if the bending modulus is known), which is computationally easier than the calculation of the local stress, which enters Equation (19).

Still, for the determination of spontaneous curvature values from Equation (18), the bending modulus needs to be known. In many practical applications of Equation (18), the bending modulus is determined with some other method [53,76–78], not relying on the local stress. However, a self-consistent way of determining $k_m$, entering Equation (18), would be the application of Equation (5), where $k_m$ is given as the second moment of the local stretching modulus $E(z)$. In Ref. [79], Campelo et al. developed a method for calculating the local stretching modulus $E(z)$ and derived the following relation:

$$E(z) = \left. \frac{\partial \left( S_L(\varepsilon, z) + P_z \right)}{\partial \varepsilon} \right|_{\varepsilon=0}, \tag{22}$$

where $S_L(\varepsilon, z)$ is the lateral part of the stress tensor at stretching $\varepsilon$, uniformly rescaled to match the reference (tensionless) configuration, assuming local incompressibility of lipid monolayers. The scaling is necessary, as due to incompressibility, the membrane thickness changes upon stretching. Equation (22) is a 3D analog of the 2D relation given by Equation (13). With $E(z)$ being found, Equation (5) can be used to determine the following elastic parameters: stretching modulus, neutral surface position, bending modulus, and spontaneous curvature.

In a subsequent elaboration of Campelo et al.'s approach, an important correction to the expression for $E(z)$, Equation (22), was found [43]:

$$E(z) = \left. \frac{\partial}{\partial \varepsilon} \left( \frac{S_L(\varepsilon, z) + P_z}{1+\varepsilon} \right) \right|_{\varepsilon=0} = \left. \frac{\partial \left( S_L(\varepsilon, z) + P_z \right)}{\partial \varepsilon} \right|_{\varepsilon=0} - \left( S_L(0, z) + P_z \right), \tag{23}$$

i.e., $E(z)$ contains not only the derivative of the lateral stress profile but also the lateral stress profile itself. If this correction is not taken into account, the systematic error of the bending modulus calculated from $E(z)$ can reach rather large values of up to 24 ± 5% [43]. Further, in Ref. [43], a more general condition of global incompressibility was taken into account rather than local incompressibility, as was in Ref. [79]. Global incompressibility implies that, while the total volume of a monolayer is preserved, the volume of any infinitesimal part of a lipid monolayer can change during deformations. Global incompressibility is thus a more general condition than local incompressibility. Due to a limited resolution of experimental techniques, it is difficult to test whether lipid membranes are locally incompressible. However, experiments show that lipid membranes are at least globally incompressible [35–38]. In addition, MD simulations demonstrate that lipid membranes are not locally incompressible [42]. In Ref. [43], with global incompressibility taken into account, it was shown that the expression for $E(z)$ transforms to:

$$E(z) = \left. \frac{\partial}{\partial \varepsilon} \left\{ \frac{\left( S_L(\varepsilon, z) + P_z \right) \beta(\varepsilon, z)}{1+\varepsilon} - P_z \frac{\partial \beta(\varepsilon, z)}{\partial \varepsilon} \right\} \right|_{\varepsilon=0}, \tag{24}$$

where $\beta(\varepsilon, z) \equiv \dfrac{dV'(\varepsilon, z)}{dV_0(\varepsilon, z)}$ is the ratio of the local volumes of the lipid material at coordinate $z$ in deformed and initial states. $\beta(\varepsilon, z)$ allows for the determination of the exact scaling map for matching the stress profiles at stretched and tensionless states [43]:

$$\zeta(\varepsilon, z) = (1+\varepsilon)^{-1} \int_0^z \beta(\varepsilon, t) dt, \tag{25}$$

where $\zeta(\varepsilon, z)$ is the $z$-dependent thickness of a monolayer at stretching $\varepsilon$ ($z = 0$ corresponds to the bilayer center). In addition, $\beta(\varepsilon, z)$ permits the determination of the local Poisson's ratio profile, $\nu(z) \equiv -\frac{1}{2}\lim_{\varepsilon \to 0}\left(\frac{\varepsilon_z}{\varepsilon}\right)$:

$$\nu(z) = \frac{1}{2(1-\gamma(z))}, \tag{26}$$

where $\gamma(z) \equiv \left.\frac{\partial S_L(0, z, P_z)}{\partial P_z}\right|_{P_z = 1\,\text{bar}} + 1$, i.e., the derivative of the lateral stress profile with respect to the isotropic ambient pressure. Local incompressibility implies that $\nu(z) = 0.5$. Simulations of coarse-grained DPPC showed considerable deviations from 0.5 in some regions of the monolayer with the largest deviation from 0.5 being $0.12 \pm 0.01$ in the glycerol region. The stress-profile approach is not the only way to determine the local Poisson's ratio. In Ref. [42], the authors directly calculated how the amount of material in a particular slab at coordinate $z$ changes upon stretching. As a definition for the material amount, the authors employed the volumes of individual atomic groups comprising lipid molecules.

In addition, in Ref. [43], a more simple procedure for the determination of elastic parameters of lipid membranes was proposed: instead of calculating $E(z)$ and then the integrals of Equation (5), it is possible to bypass the calculation of $E(z)$. Actually, by definition (see Equation (3)):

$$E(z) = \left.\frac{\partial^2 w_{3D}^{\text{mono}}(\varepsilon, z)}{\partial \varepsilon^2}\right|_{\varepsilon = 0} = \left.\frac{\partial}{\partial \varepsilon}\sigma(\varepsilon, z)\right|_{\varepsilon = 0}, \tag{27}$$

where $\sigma(\varepsilon, z)$ is called the local tension profile. Equation (27) allows for replacing the integral and derivative signs after the substitution into Equation (5). Thus, the expression for the bending modulus, for example, becomes [43]:

$$k_m = \left.\frac{\partial}{\partial \varepsilon}\int_{m_0} \sigma(\varepsilon, z)(z - z_0)^2\, dz\right|_{\varepsilon = 0}, \tag{28}$$

which is computationally easier than first calculating $E(z)$, as only one curve fitting of simulation data is required for employing Equation (28). Note that the integral expression on the right-hand side of Equation (28) at $\varepsilon = 0$ coincides with the Gaussian curvature modulus; see Equation (5):

$$\bar{k}_m = \int_{m_0} \sigma_0(z)(z - z_0)^2\, dz, \tag{29}$$

where $\sigma_0(z) = \sigma(0, z)$ by definition; see Equations (3) and (27). However, at $\varepsilon \neq 0$, the integral on the right-hand side of Equation (28) does not equal $\bar{k}_m(\varepsilon)$, an $\varepsilon$-dependent Gaussian curvature modulus. Nevertheless, for the quadratic incompressible model of a lipid monolayer, i.e., within the assumptions of quadratic elastic energy and local incompressibility, it is possible to derive that [43]:

$$k_m = 2\bar{k}_m(0) + \left.\frac{d}{d\varepsilon}\bar{k}_m(\varepsilon)\right|_{\varepsilon = 0}. \tag{30}$$

Equation (30) shows that $\bar{k}_m(\varepsilon)$ contains information about the bending modulus. However, the results of Ref. [43] show a considerable deviation from Equation (30): Equation (30) predicts the bending modulus value two times less than the actual value, which was attributed to the inapplicability of the quadratic incompressible model.

Although Equation (29) is frequently employed in MD simulations for the determination of the Gaussian curvature modulus, the resulting values are often positive [78,80], which contradicts the stability condition of the Helfrich model, $-2k \leq \bar{k} \leq 0$ [52]. However, in Ref. [14], it was shown that even if $-2k \leq \bar{k} \leq 0$ holds, the Gaussian curvature term in the elastic energy still causes instability if the tilt deformation mode is taken into account. It was, therefore, suggested that the elastic energy should be written together with the Gaussian curvature squared or omitted from the elastic energy to satisfy the stability conditions [14]. In Refs. [40,44], it was proposed that, depending on the lateral fluidity assumption, the contribution from the local lateral shear modulus may additionally enter the expression for the Gaussian curvature modulus; see Section 2.2.2. for details.

The local stress methods have a drawback regarding ambiguity in the definition of the local stress. The problem is that a force decomposition into pairwise contributions is required for the calculation of the local stress [48,81,82]. This problem concerns multibody potentials, i.e., potentials that depend on the positions of more than two particles. Such multibody potentials are employed in MD simulations of lipids. For example, four-body potentials are used for the description of the rotations around the double bonds of hydrocarbon chains [48]. Currently, the so-called covariant central force decomposition (cCFD) is mostly employed [82]. Unlike the Goetz–Lipowsky decomposition [66], cCFD is consistent with Newton's laws of motion and, in particular, with the conservation of angular momentum [82]. In Ref. [83], however, it was shown that it is possible to construct another force decomposing, a force center decomposition (FCD), which is also consistent with Newton's laws of motion. Moreover, it was shown that the elastic parameters can depend on the force decomposition employed: although the first moment of the stress profile remains invariant, the second moment is sensitive to the choice of the force decomposition [83,84].

In summary, within the equilibrium force methods, the planar configuration of lipid membranes allows for the determination of all major elastic parameters: bending, stretching and bulk moduli, spontaneous curvature, and the local Poisson's ratio. The advantage of the planar configuration is that it has a rather simple setup: a square planar lipid membrane with periodic boundary conditions. However, one should be cautious with the system size, as for large systems, the effect of thermal undulations becomes important and may impact theoretical predictions. Attempts exist to explicitly take into account these undulations, but at short wavelengths, these undulations become hardly predictable (see Section 4.3). Although the macroscopic stress can be strictly defined, there is still no consensus on the definition of the local stress: the local stress is not uniquely defined, as there are many definitions that all average to the same macroscopic stress, and there is no criterion that determines whether a particular decomposition is more fundamental than other ones.

*3.2. Tubular Membranes*

3.2.1. Bilayer Tethers

A cylindrical geometry of lipid membrane tubes (see Figure 3B), also called tethered, is another useful configuration for the determination of lipid membrane elastic parameters. Lipid tubes are also widely used in experiments for the determination of elastic parameters [31,85,86]. In this setup, a cylindrical lipid membrane is created [87–89]. From Equation (1), it follows that the stress–strain relation in this configuration is [52]:

$$f = 2\pi k \left( \frac{1}{R} - K_0 \right), \tag{31}$$

where $f$ is the force along the tube axis, and $R$ is the tube radius. For lipid bilayers, $R$ is usually chosen to be the distance from the tube axis to the midsurface between two monolayers [87]. Employing Equation (31), it is possible to determine both the bending modulus $k$ and spontaneous curvature $K_0$ by measuring the axial force and tube radius [87–89]. Equation (31) is usually used to determine elastic parameters of symmetric lipid bilayers, which implies that $K_0 = 0$. The area per lipid should be equalized in both monolayers before the measurements. The same applies to the internal (inside a tube) and external (outside a tube) pressures. However, the self-equilibration of the area per lipid takes a rather long time, as the lipid flip-flop rate is usually slow. To facilitate this process, small pores oriented perpendicular to the tube axis can be intentionally introduced [89].

As in the case of planar bilayers, an undulation correction exists for Equation (31) [90]:

$$f = \frac{2\pi k}{R}\left(1 - \frac{k_B T}{2\pi^2 k} R^2 \Lambda^2 \right), \quad (32)$$

where $K_0$ is assumed to be zero, and $\Lambda$ is the cutoff wave vector. This correction becomes important at large $R$; see Figure 3 of Ref. [90].

In principle, it is possible to apply the method of tubular bilayers not only to single-component membranes but also to multicomponent ones. However, as the outer and inner monolayers have different curvature radii, the equilibrium composition of these monolayers is different due to the composition–curvature coupling [91]. Therefore, the bending modulus of individual monolayers can be different, which complicates the determination of intrinsic bending moduli of constituent monolayers.

Overall, by simulating tubular bilayers, it is possible to determine the bilayer bending modulus and spontaneous curvature. In addition, if a membrane consists of only one lipid type, the monolayer bending modulus can be inferred as half of the bilayer bending modulus. However, this reasoning does not work for the monolayer spontaneous curvature. In addition, the determination of elastic parameters of multicomponent monolayers is complicated due to the composition–curvature coupling effect.

3.2.2. Inverted Hexagonal Phase

Bilayer tethers are not the only tubular structure employed for the determination of elastic parameters. Another tubular structure frequently used both in experiments and MD simulations is the inverted hexagonal (H$_{II}$) phase. Some lipids, such as dioleoylphosphatidylethanolamine (DOPE), which has a large negative spontaneous curvature, tend to form a so-called inverted hexagonal phase in aqueous solutions [92]. This phase consists of monolayer cylinders (see Figure 3C) located parallel to each other in a hexagonal lattice with lipid heads being oriented inside the cylinders. The H$_{II}$ phase is frequently employed for the determination of intrinsic spontaneous curvature of lipids in experiments [49,50,93]. For this, a binary lipid mixture is employed, the H$_{II}$-forming lipid of which is referred to as 'host' and the other is called 'guest'. From Equation (31), it follows that when $f = 0$:

$$K_0 = \frac{1}{R}, \quad (33)$$

where $K_0$ is the spontaneous curvature of the binary mixture, and $R$ is the radius of a cylinder. Within the additivity assumption, $K_0$ can be expressed as:

$$K_0 = (1-x)K_0^h + xK_0^g, \quad (34)$$

where $K_0^h$ and $K_0^g$ are the intrinsic spontaneous curvatures of the host and guest lipids, respectively, while $x = \dfrac{N_g}{N_g + N_h}$ is the concentration of the guest lipid, where $N_g$ and $N_h$ are the number of the guest and host lipids, respectively. The additivity assumption of Equation (34) does not always hold [43,75]. The combination of Equation (33) and Equation (34) yields:

$$\frac{1}{R} = K_0^h + x\left(K_0^g - K_0^h\right), \tag{35}$$

which allows for the determination of $K_0^g$ from the slope of $\frac{1}{R}$ as a function of $x$. The measurement of $R$ requires the determination of the reference surface. Experiments [93] as well as MD simulations [94] show that the $H_{II}$ monolayers contain a surface, at which the area per lipid remains constant upon deformations. This surface is usually referred to as the pivotal surface. Theoretical estimations show that the spontaneous curvature radii with respect to the neutral and pivotal surfaces are related as [93]:

$$R_p = R_n \left(\frac{1+\gamma}{1-\gamma}\right)^{1/2}, \tag{36}$$

where $R_p$, $R_n$ are the radii of the pivotal and neutral surfaces, respectively, and $\gamma = \frac{k_m}{k_{A,m} R_n^2}$. The typical values of $\gamma$ are $\leq 10\%$ [93], which means that the neutral and pivotal surfaces almost coincide. Therefore, although Equation (33) is valid only for the pivotal surface [93], the small value of $\gamma$ allows for neglecting the difference between the neutral surface and pivotal surface for the determination of the spontaneous curvature. MD simulations [94] and experiments [49] show that the pivotal surface lies in the region of the lipid backbone. While the spontaneous curvatures relative to the neutral and pivotal surfaces do not differ much, the difference between the bending moduli relative to these surfaces can be as large as 40% [93].

Equation (33) is valid when the pressure difference between the interior and exterior of the $H_{II}$ cylinders, $\Delta p$, is equal to zero. If the pressure difference is nonzero, the following equation is satisfied [93,94]:

$$2k_{m,p}\left(\frac{1}{R_p} - \frac{1}{R_{0,p}}\right) = -R_p^2 \Delta p, \tag{37}$$

where $k_{m,p}$ is the bending modulus relative to the pivotal surface and $R_{0,p}$ is the curvature radius of the pivotal surface at $\Delta p = 0$. Equation (37) reflects the force balance between the bending stress and the force due to the pressure difference. In Ref. [94], Equation (37) was employed for the determination of $k_{m,p}$ and $R_{0,p}$ by simultaneously measuring $R_p$ and $\Delta p$ at a given value of the lipid/water ratio, by analogy with experiments [93]. The standard error in the obtained value of $k_{m,p}$, however, was quite large ($\approx 45\%$) due to difficulties in the calculation of $\Delta p$, which is a rather noisy quantity. In Ref. [95], an alternative approach was proposed, which instead of measuring $\Delta p$, involves the measurement of the pressure along ($P_{\parallel}$) and perpendicular ($P_{\perp}$) to the cylinder axis. It was shown that it is possible to obtain analytical expressions for $P_{\parallel}$ and $P_{\perp}$, given by Equations (12) and (13) of Ref. [95], which are too bulky to be presented here. Then, performing the non-linear regression of $P_{\parallel}$ and $P_{\perp}$ obtained at constant volume and different cylinder lengths allows for the determination of the spontaneous curvatures of guest and host lipids as well the bending and stretching moduli. In Ref. [95], the bending and stretching moduli were assumed to be independent of the lipid composition, which is a rather significant simplification and may not hold for binary mixtures of, for example, lipids with different values of the bending moduli as, for instance, lipids with a different degree of unsaturation [69].

*3.3. Buckling*

Buckling is the bulging of thin elastic materials out of a plane as a result of a significant compression in the lateral direction; see Figure 3D. Such bulging also occurs in fluid films [96] and lipid membranes [71,97,98]. As a method for the determination of lipid membrane elastic parameters, the buckling procedure was introduced in Refs. [71,96–98]. In this approach, the simulation starts with a planar lipid membrane along the $xy$-plane in a simulation box with side lengths $(L_x, L_y, L_z)$ with $L_x$ being several times larger than $L_y$. Then, a compressive force is applied along the $x$-direction with $L_y$ kept fixed to create a buckled configuration at some fixed value of $\gamma \equiv \frac{L - L_x}{L}$, where $L$ and $L_x$ are the side lengths of a simulation box along the $x$-direction of the initial and deformed state, respectively. The value of the critical lateral tension, $\sigma_c$, at which the transition to the buckling regime occurs can be estimated from the divergence of the shape fluctuation spectrum at the longest wavelength [71,97]. For planar lipid membranes, this spectrum is $\sim (kq^4 + \sigma q^2)^{-1}$ [30], where $q$ is the magnitude of the wave vector. For the smallest value of $q_{\min} = \frac{2\pi}{L}$, the divergence occurs at [71,97]:

$$\sigma_c = -kq_{\min}^2 = -\frac{4\pi^2 k}{L^2}. \tag{38}$$

From Equation (38), it follows that the critical tension is smaller for larger systems. The linear relationship between the critical tension and $1/L^2$ as predicted by Equation (38) was confirmed in Refs. [71,97], which allowed for the determination of the bending modulus $k$ as a slope of the critical tension as a function of $4\pi^2 / L^2$.

The analysis of the buckled configuration was further elaborated in Refs. [96,98]. It was shown that the stress–strain relation, which follows from the Helfrich Hamiltonian, of the buckled configuration satisfies the following relations [98]:

$$f_x = k \left(\frac{2\pi}{L}\right)^2 \left(1 + \frac{1}{2}\gamma + \frac{9}{32}\gamma^2 + ...\right), \tag{39a}$$

$$f_y = k \frac{(2\pi)^2}{A} \frac{L_y}{L_x} \left(1 - \frac{5}{2}\gamma - \frac{23}{32}\gamma^2 + ...\right), \tag{39b}$$

where $A = LL_x$, and $f_x, f_y$ are the stresses (forces per unit length) along the $x$- and $y$-directions, respectively. Thus, by fitting simulation data by the functions given in Equation (39a,b), one can determine the bending modulus $k$. However, a rather large system is required for the implementation of the buckling method ($L \approx 40$ nm, $L_y \approx 7$ nm [98]), which leads to the necessity to make undulation corrections to Equation (39a,b) [98]. Although the correction for $f_x$ is negligibly small, the correction for $f_y$ can be up to 10% [98]:

$$\delta f_y = -k_B T \frac{L}{L_y a} \left[1 + \frac{3L_y a}{2L^2}\left(1 - \frac{11}{8}\gamma + ...\right)\right], \tag{40}$$

where $a$ indicates the microscopic cutoff length of membrane undulations. With this correction taken into account, both Equation (39a) and (39b) yield consistent results for the bending modulus.

Apart from fluid symmetric bilayers, the buckling approach can also be applied to bilayers in the gel phase [99] and asymmetric bilayers [100,101]. The theoretical analysis of these systems, however, requires significant corrections to Equation (39a,b) due to the softening effect: the bending modulus of these systems decreases with curvature. This softening effect can be incorporated into the elastic energy in the following phenomenological way [99]:

$$W_S = kl^{-2}\left(\sqrt{1+l^2K^2}-1\right), \tag{41}$$

where $l$ is some curvature scaling factor. With this elastic energy, the stress–strain relation of Equation (39a) transforms to [99]:

$$f_x = k\left(\frac{2\pi}{L}\right)^2\left(1+\frac{1}{2}(1-3\delta^2)\gamma+...\right), \tag{42}$$

where $\delta = 2\pi l / L$ is a softening parameter, which enters the stress–strain relation as an additional fitting parameter. In Ref. [101], it was shown that the asymmetry in the lipid composition can lead to the phase transition in the compressed leaflet of a buckle, which leads to the increase in the bending modulus $k$ and softening parameter $\delta$. Asymmetric bilayers are of particular interest, as several biological membranes, such as membranes of endosomes or plasma membranes, are usually asymmetric in lipid composition [102]. It was also shown that such a phase transition occurs more readily in larger systems [101], which indicates that this effect may occur in living cells, influencing the elastic properties of their membranes.

The buckling method can also be applied for the determination of the tilt modulus [103]. The Euler–Lagrange equation following from the HK Hamiltonian, Equation (4), for the tilt field can be written as:

$$T''(s) - l^2 T(s) = -K'(s), \tag{43}$$

where $s$ is a natural parametrization parameter for the buckled configuration and $l = \sqrt{\frac{k_m}{k_t}}$. Within the approximation $|T''| \ll \left|\frac{T}{l^2}\right|$, it is possible to derive the following equation for the tilt modulus [103]:

$$k_t = k_m\left(\frac{2\pi^2}{L}\right)\frac{1+\frac{1}{2}\gamma+\frac{9}{32}\gamma^2+...}{z_0'(\xi)}, \tag{44}$$

where $z_0(\xi)$ is the pivotal plane position, determined with respect to the reference atom at a distance $\xi$ from the midsurface. The pivotal plane is a special plane parallel to the monolayer surface, at which the area per lipid remains the same during deformations, see also Section 3.2.2. It was shown that the location of this plane can be determined from the difference in the number of lipids between the opposing leaflets in curved geometries such as spherical and cylindrical [104]. The determination of the pivotal plane is also possible in the buckled configuration by looking at the leaflet difference at different membrane sections [104]. In this framework, the number of lipids in a particular section depends on the choice of a reference atom, the coordinates of which define the position of lipids relative to a particular membrane section. This leads to a $\xi$-dependent position of the pivotal plane, $z_0(\xi)$, which enters Equation (44). Equation (44) follows from the HK Hamiltonian, which does not take into account the tilt–curvature coupling term of a more general Hamiltonian, Equation (6), which was subsequently discovered, and the effect of this tilt–curvature coupling term on Equation (44) was not yet analyzed.

Currently, the theoretical framework for the buckling protocol exists only for single-component lipid bilayers [96,98]. The consideration of multi-component lipid bilayers is complicated by the composition–curvature coupling [86,105]: a lateral redistribution of lipids to regions of different curvature. This effect leads to an inhomogeneous concentration of lipid components and thus to the composition-dependent elastic parameters: $k_m \to k_m(\mathbf{r})$, $K_0 \to K_0(\mathbf{r})$, etc., where $\mathbf{r}$ is the position vector. Formally, the buckling method can be applied to multicomponent lipid membranes [106]. The bending modulus obtained in this way reflects some mean value of the overall system. However, this value is buckling-specific, as in general, $k_m(\mathbf{r})$ is different in other curved geometries. Position-dependent elastic parameters can be referred to as *intrinsic* ones, as they reflect the parameters of a specific lipid composition at $\mathbf{r}$. To avoid the composition–curvature coupling effect, intrinsic elastic parameters of multicomponent lipid membranes can be determined in geometries with fixed curvatures such as planar or cylindrical bilayers. The constancy of the extrinsic curvatures assures that the lipid composition remains homogenous and the composition–curvature coupling effect is not involved.

In summary, the buckling procedure is a rather powerful technique, which permits the determination of not only the bending modulus but also the tilt modulus and pivotal plane position. However, currently, the theoretical framework for the buckling approach exists only for single-component lipid membranes: the application to multicomponent membranes is complicated by the composition–curvature coupling effect. In addition, the buckling procedure does not provide the spontaneous curvature values of constituent monolayers.

*3.4. Sinusoidal Bilayers*

In this approach [107], a lipid bilayer is sinusoidally deformed by two cylindrical guiding potentials, see Figure 3E. The potentials are given as cylindrical rigid wall potentials that interact with hydrophobic beads of a considered membrane. The cylinders are located on opposite sides of a membrane parallel to each other. For a planar membrane spanning from 0 to $L$ along the $x$-axis, the axes of the cylinders are located at $L/4$ and $3L/4$. The coordinates of the points where the cylinders touch a membrane are chosen to be equal up to a sign and are denoted by $z_m$ and $-z_m$. From the Euler–Lagrange equation, following from the Helfrich Hamiltonian under the assumption $z_m/L \ll 1$, it follows that the force $F_z$ acting on the cylinders equals:

$$F_z = 96k\frac{L_y z_m}{L^3}, \tag{45}$$

where $L_y$ is the side length of the simulation box along the $y$-axis. Equation (45) is valid only for $z_m/L \ll 1$. The exact stress–strain relation does not have a closed-form analytical solution and can be found numerically. Then, the simulation data can be fitted to this numerical solution to determine the bending modulus $k$. In this approach, one should be cautious with the definition of $z_m$. It was shown that $z_m = 0$ differs for membranes with different $L$ due to thermal undulations [107]. In practice, $z_m = 0$ is chosen to be a limit at $L \to 0$ [107]. The approach of Ref. [107] resembles that of buckling [96,98]. However, a lack of an analytical solution to the boundary value problem for the configuration of Ref. [107] makes the buckling procedure more suitable for applications. In addition, there is no need to introduce any external potentials in the buckling approach wherein lipid bilayers naturally buckle, staying within the periodic boundary conditions.

*3.5. Spontaneously Curved State*

In Ref. [108], an approach was suggested for the determination of the spontaneous curvature of asymmetric bilayer membranes. In this work, lipid bilayers consisting of two lipid types, A and B, were considered. Each leaflet was divided into two equal parts: the first part is composed of lipid A of ratio $\rho$ and lipid B of ratio $1-\rho$, while in the second part, the ratios are $1-\rho$ and $\rho$ for lipid A and B, respectively. The opposing leaflet is constructed similarly, such that the lipid bilayer is in an antisymmetric configuration. If the diffusion of lipids is artificially prevented, from the Helfrich Hamiltonian, it follows that such a lipid bilayer relaxes to a sine-like shape with the extrinsic curvature equal to the spontaneous curvature [108]. Coarse-grained MD simulations showed that in the equi-librium state, the membrane divides into two regions with approximately constant curvature of opposite signs. At $\rho = 0.5$, the bilayer spontaneous curvature appeared to be a linear function of the head bead diameter difference between lipids A and B. In addition, at a constant head bead diameter difference, the spontaneous curvature showed a linear dependence on $\rho$. The approach of Ref. [108] considers only the bilayer spontaneous curvature, not the monolayer spontaneous curvature. Thus far, this method has been applied only to coarse-grained lipids represented by two hydrophobic and one hydrophilic bead. Note that a modification of a force field is required for preventing the diffusion of lipids. This modification is assumed to not influence spontaneous curvature values, as this modification concerns only fluidity properties of a membrane.

*3.6. Collective-Variable Methods*

In this approach, a bias potential $V_{bias}\left[\xi(\mathbf{s})\right]$ is introduced that depends on some collective variable $\xi(\mathbf{s})$, where $\mathbf{s} \equiv \mathbf{s}(\{\mathbf{r}_i\})$ with $\{\mathbf{r}_i\}$ being the collection of particles' positions of a considered system. $V_{bias}\left[\xi(\mathbf{s})\right]$ is usually set by a harmonic potential, $V_{bias}(\xi) = \frac{\chi}{2}(\xi - \xi_0)^2$, which permits the determination of the free energy $F(\xi_0)$ employing the umbrella sampling [109]. Collective variables are frequently employed for reshaping lipid membranes to various geometries with nonzero curvatures [91,110–114]. However, it is usually difficult to single out all elastic contributions corresponding to $F(\xi_0)$, as in general, each value of $\xi$ can be characterized by the presence of different deformation modes.

In Ref. [91], the authors employed the local density of hydrophobic beads as a collective variable. The local density values were sampled from the unbending of a semi-cylindrical membrane; see Figure 3F. At each snapshot of the unbending trajectory, the midsurface of a lipid bilayer can be well approximated by a parabolic profile, the elastic energy of which can be calculated from the Helfrich Hamiltonian. The bending modulus can thus be obtained from the comparison of this elastic energy with the calculated free energy along the unbending trajectory. To calculate the elastic energy, a truncation of the midsurface at the membrane edges is required. The choice of this truncation, however, is not strictly defined, which can contribute to the uncertainty in a measured bending modulus [91].

The same process, i.e., unbending of half-cylinders, was employed in Ref. [115]. However, the collective variable was defined differently from Ref. [91]. In Ref. [115], the authors employed the permutation reduction scheme and principal component analysis to identify a reaction coordinate (collective variable). The simulation results showed that the bending modulus increases with the extrinsic curvature, i.e., the Helfrich Hamiltonian fails at small radii of extrinsic curvatures, $R \leq 10$ nm [111]. As in Ref. [91], in Ref. [115], the curvature elastic energy was determined via the integration over the midsurface of a bilayer. It was later shown that the interpretation of the simulation data of Ref. [111] depends on the choice of the reference surface used for the calculation of the elastic energy: it was argued that a monolayerwise application of the Helfrich Hamiltonian is more appropriate for this problem and restores the constancy of the curvature elastic energy [116].

In Refs. [80,117], a spherical deformation was applied to an initially flat bilayer patch. To accomplish this, an artificial external potential was applied to the outer tail-beads of all lipids to create a spherical cap of a fixed mean curvature (half of the extrinsic curvature) $c$. Unlike previously described methods, umbrella sampling was not employed. Instead, the system was freed to relax to an equilibrium state, which is either a closed bilayer vesical or planar bilayer patch. It can be shown that the elastic energy of these spherical caps relative to the planar state, $\Delta W$, can be written as:

$$\frac{\Delta W(x,\xi)}{4\pi(2k+\bar{k})} = x + \xi\left[\sqrt{1-x} - 1\right], \tag{46}$$

where $x = (Rc)^2$ and $\xi = \dfrac{\gamma R}{2k + \bar{k}}$ with R corresponding to the radius of the closed vesicle and $\gamma$ being the edge tension of the cap circumference, which is taken into account in addition to Equation (1). For each value of x, it is possible to determine the probability of the cap relaxation to the closed vesicle, which corresponds to x = 1, for which an analytical expression can be determined [117]. The approach of Refs. [80,117] was originally developed for the determination of the Gaussian curvature modulus $\bar{k}$, as a spherical configuration has two nonzero principal curvatures, unlike planar or cylindrical configurations. Therefore, the bending modulus k, which enters Equation (46), was determined independently by employing lipid tubes (see Section 3.2). The edge tension was also determined independently by employing planar bilayers with edges [117]. For a Cooke model [54,55], it was shown that $\dfrac{\bar{k}}{k}$ lies within the range −0.95 ± 0.1 [117]. However, $\bar{k}$ measured from the stress profile appeared to disagree with that of Equation (46), which was attributed to the ambiguity in the definition of the local stress [117].

Currently, the collective-variable approaches are limited to the determination of either the bilayer bending modulus or Gaussian curvature modulus. In addition, the considered collective-variable approaches were applied only for single-component membranes. The application of the collective-variable methods to multicomponent membranes can be complicated by the composition–curvature coupling. For example, during the relaxation of half-cylinders as in Refs. [91,111], the shape soon becomes parabolic, which does not have a constant curvature and can involve the redistribution of lipids to regions of different curvatures in multicomponent membranes, complicating the determination of intrinsic elastic parameters.

## 4. Fluctuation-Based Methods

In the previous section, the methods that rely on the average values of stresses and strains were considered. However, any macroscopic physical system always undergoes thermal fluctuations: random deviations of system parameters from their average values. Lipid membranes are not an exception, and in this section, MD approaches that analyze these fluctuations to infer the elastic parameters of lipid membranes are reviewed.

### 4.1. Fluctuations of Surface Area

In Section 3.1.1, it was already shown that the stretching modulus of lipid membranes can be obtained from the dependence of the membrane area on the lateral tension. The latter approach relies on the average values of the lateral tension and membrane area. However, the membrane area constantly fluctuates around the average value. The characteristics of these fluctuations can be derived from the rules of statistical mechanics. The elastic energy of stretching of a lipid bilayer $\alpha$ equals $A_0 \dfrac{1}{2} k_A \alpha^2$, where $k_A$ is the bilayer stretching modulus and $A_0$ is the average area of a lipid bilayer in the reference tensionless state. According to the Boltzmann distribution, the probability of a state with stretching $\alpha$ is proportional to $\exp\left(-\dfrac{A_0 k_A \alpha^2}{2 k_B T}\right)$, from which it follows that [68]:

$$k_A = \frac{k_B T}{A_0 \langle \alpha^2 \rangle} = \frac{k_B T \langle A \rangle}{\langle \delta A^2 \rangle}, \qquad (47)$$

where the triangular brackets indicate the averaging over time, A is the total membrane area and $\langle \delta A^2 \rangle = \langle A^2 \rangle - \langle A \rangle^2$. As with the stress–strain relation of Equation (11), Equation (47) applies only to sufficiently small systems, for which the contribution of thermal undulations to $k_A$ can be neglected.

In Equation (47), $k_A$ refers to the stretching modulus of a lipid bilayer. For symmetric bilayers, the equality $k_A = 2k_{A,m}$ holds. For asymmetric bilayers, it can be shown that [118]:

$$\frac{1}{k_A} = \frac{1}{2}\left(\frac{1}{k_{A,m}^u} + \frac{1}{k_{A,m}^l}\right), \tag{48}$$

where $k_{A,m}^u$, $k_{A,m}^l$ are the local monolayer stretching moduli of the upper and lower monolayers, respectively. In Ref. [118], it was proposed that the local monolayer stretching moduli should be distinguished from the global monolayer stretching moduli, which are equal to $\frac{k_A}{2}$ in symmetric bilayers. Note that Equation (47) does not apply to the determination of $k_{A,m}^u$ and $k_{A,m}^l$, as both $\langle A \rangle$ and $\langle \delta A^2 \rangle$ are identically equal in both monolayers due to the periodic boundary conditions. Therefore, a different method was proposed for the determination of $k_{A,m}^u$ and $k_{A,m}^l$ [118]: $k_{A,m}^u$, $k_{A,m}^l$ can be determined from the fluctuations of the local thickness of monolayers. The local thickness appears more suitable for MD calculations than the local area [118]. For each leaflet, the volume incompressibility permits expressing the elastic energy of stretching through the leaflet thickness [118]:

$$E^L = \frac{1}{2} k_{A,m}^L a_0^L \left(\frac{t_0^L - t^L}{t^L}\right)^2, \tag{49}$$

where $k_{A,m}^L$ is the stretching modulus of a considered leaflet, $t_0^L$ is the average leaflet thickness, $a_0^L$ and $t^L$ are the average area and instantaneous thickness of some local region of a leaflet, respectively. Applying the Boltzmann distribution to the energy levels of Equation (49), one obtains [118]:

$$-\frac{2k_\text{B}T}{a_0^L} \ln p\left(\frac{t_0^L - t^L}{t^L}\right) = k_{A,m}^L \left(\frac{t_0^L - t^L}{t^L}\right)^2 + C, \tag{50}$$

where $p$ is the probability. $k_{A,m}^L$ can also be calculated from the equipartition theorem (analogous to Equation (47)). However, Equation (50) has the advantage of employing the truncation for the large values of $\frac{t_0^L - t^L}{t^L}$, which does not follow the quadratic elastic regime. It was shown that the most relevant surface to employ for the definition of the local thickness lies approximately in the middle of the hydrocarbon chains. The local thickness, thus, is the distance between the terminal methyl carbons and this relevant surface. However, if high concentrations of cholesterol molecules are considered, this definition requires corrections to be made to the measured values of monolayer stretching moduli [118].

Recall that $k_{A,m}$ is equal to the integral of the local stretching modulus $E(z)$ over the monolayer thickness, $k_{A,m} = \int E(z)\,dz$. When the fluctuations are analyzed with the help of Equation (47) or Equation (50), the information about $E(z)$ remains hidden. In Section 3.1.2, the equilibrium force approach for the determination of $E(z)$ was considered. The question is whether $E(z)$ can also be determined from fluctuations. In Refs. [119,120], the authors equated $E(z)$ with the fluctuations of the depth-dependent area occupied by lipid molecules, $A_{occ}(z)$:

$$E(z) = \frac{k_B T \langle A_{occ}(z) \rangle}{\langle \delta A_{occ}^2(z) \rangle}. \tag{51}$$

$A_{occ}(z)$ was determined by dividing the simulation box into the grid cells with side lengths of approximately 0.03 × 0.03 × 0.01 nm along the *x*-, *y*- and *z*-axis, respectively. A grid point was assumed to be occupied by a lipid molecule if any atom of this lipid lies within the van der Waals radius from the grid point. Then, $A_{occ}(z)$ was calculated as a sum of the areas of the occupied grid cells. Thus, Equation (51) is analogous to Equation (47) with $A$ replaced to $A_{occ}(z)$ and enables the calculation of the depth-dependent stretching modulus $E(z)$.

*4.2. Fluctuations of Volume*

The bulk modulus $k_V$ expressed through average values of volume and pressure is given by Equation (2). By analogy with the stretching modulus, it is possible to derive a fluctuation-based expression for $k_V$. The elastic energy density per unit volume is:

$$w_V = \frac{1}{2} k_V \alpha_V^2, \tag{52}$$

where $\alpha_V = \frac{V - V_0}{V_0}$ with $V$, $V_0$ being the deformed and initial volume, respectively. The relation between $k_V$ and volume fluctuations is thus [121,122]:

$$k_V = \frac{k_B T \langle V \rangle}{\langle \delta V^2 \rangle}, \tag{53}$$

where $V$ is the total volume of a lipid membrane and $\langle \delta V^2 \rangle = \langle V^2 \rangle - \langle V \rangle^2$. Equation (53) is similar to Equation (47): only the averaging values differ. Equation (53) was employed in Ref. [121] for the determination of the bulk modulus of atomistic DPPC bilayers with the volume boundary defined at the position of phosphorus atoms, and the obtained value of $k_V$ was 0.6 GPA. DPPC bilayers were also studied in Ref. [122], and the value of 1.5 GPa for the bulk modulus was reported. However, in Ref. [122], Equation (53) was applied to the simulation box as a whole, i.e., with water molecules included. Disentangling the contributions to the bulk modulus from lipids and water leads to a slightly smaller value of 1.3 GPa [42].

*4.3. Fluctuations of Shape*

The shape of lipid membranes, as well as surface area and volume, constantly changes due to thermal motion. Like the area and surface fluctuations, the shape fluctuations can be described using the rules of statistical mechanics. Let us consider a planar square lipid membrane of area $A$ in the *xy*-plane with periodic boundary conditions. From the Helfrich Hamiltonian, it follows that the fluctuations of the shape $h(x,y)$ of this membrane satisfy the following equation [30]:

$$\langle |h_\mathbf{q}|^2 \rangle = \frac{k_B T}{k q^4 + \sigma q^2}, \tag{54}$$

where $\mathbf{q} = \frac{2\pi}{\sqrt{A}} \begin{pmatrix} n_x \\ n_y \end{pmatrix}$, $n_x, n_y \in \mathbb{Z}$ is the wave vector, $q = |\mathbf{q}|$ and $h = \frac{1}{\sqrt{A}} \sum_\mathbf{q} h_\mathbf{q} e^{i\mathbf{q}\cdot\mathbf{r}}$ with $\mathbf{r}$ being the position vector. Equation (54) follows from the equipartition theorem applied to the Helfrich Hamiltonian written in the Fourier space. Equation (54) is a theoretical result for infinitely thin fluid films. In MD simulations, a reference surface should be chosen for the calculation of $h_{\mathbf{q}}$. The first application of Equation (54) to MD simulations of planar bilayers was introduced in Ref. [67]. The reference surface was chosen to be the midsurface between the two monolayers. Although the simulations were performed at $\sigma = 0$, the fluctuation spectrum showed two regimes: $\sim 1/q^4$ at low $q$ and $\sim 1/q^2$ at high $q$. The transition between the two regimes occurs at a wavelength comparable with the monolayer thickness, $\approx 2.5$ nm. The $1/q^2$ dependence at high $q$ was attributed to the protrusion modes, i.e., relative displacements of individual lipid molecules, which roughen the shape of the reference surface [67]. It was later shown that the discrepancy from Equation (54) at high $q$ values can be explained by the tilt degree of freedom, which, if taken into account, leads to the following fluctuation spectrum for lipid monolayers [123] at $\sigma = 0$:

$$\left\langle \left| h_{\mathbf{q}} \right|^2 \right\rangle = \frac{k_B T}{k_m q^4 + k_t q^2}, \tag{55}$$

which is the same as Equation (54) if $\sigma$ is replaced by the tilt modulus $k_t$. The deviation from Equation (54) at high $q$ values, however, is realized not only for 3D lipid membranes but also for 2D meshless membranes [88], which is again attributed to protrusions [88]. It was shown that a proper averaging procedure over grid cells can recover the theoretical result of Equation (54) at high $q$ values. In general, the protrusions are coupled with the bending deformation mode. In Refs. [124,125], it was shown that the protrusions are disentangled from the coupled undulatory mode $\left\langle \left| h_{\mathbf{q}}^{CU} \right|^2 \right\rangle \equiv \left\langle h_{\mathbf{q}}^{low} h_{\mathbf{q}}^{up*} \right\rangle$, where $h_{\mathbf{q}}^{low}$ and $h_{\mathbf{q}}^{up}$ are the Fourier amplitudes of the shape of the lower and upper monolayer, respectively, which enables a more accurate determination of the bending modulus both from the fluctuations of the membrane shape [124] and density of phosphorus atoms [125]. In practice, the bending modulus $k$ is usually determined by sampling at low $q$ values to avoid protrusions. The sampling at low $q$ values is however complicated by a slow relaxation rate of small $q$ deformation modes, which scales as $1/q^3$ [5,98] due to a solvent viscosity. To enhance the sampling at low $q$ values, methods have been proposed that use umbrella enhancement of long-wavelength deformations [126,127].

In Refs. [128–130], it was shown that the tension appearing in the Helfrich Hamiltonian (also called intrinsic tension), Equation (1), and the mechanical tension (also called frame tension), which can be obtained from the macroscopic stress tensor, do not coincide due to undulations, which might explain the discrepancy at high $q$ values between the observed spectrum and the theoretical prediction of Equation (54). However, a rigorous theoretical analysis of the Helfrich Hamiltonian shows that the coefficient of $q^2$ in the fluctuation spectrum should be the frame tension, not the intrinsic tension [131–133], implying that microscopic degrees of freedom are likely responsible for the observed discrepancy rather than the difference between the intrinsic and frame tensions [133].

The application of Equation (54) requires the choice of the reference surface to be made. Actually, lipid bilayers consist of two monolayers, and each monolayer has its reference surface. A comprehensive study of systematic errors in fluctuation-based methods shows that the ambiguity in the choice of the reference surface can lead to systematic errors in the bending modulus, being as large as 75% (according to row 1 of Figure 5 in Ref. [134]). It can be shown, however, that some choices are more reasonable than others. In particular, the head group region appears to be more favorable than the tails region, as the lipid tails tend to undergo quite large fluctuations [134].

In principle, Equation (54) can be applied for the bending modulus determination not only to single-component lipid bilayers but also to multi-component bilayers. However, the composition–curvature coupling can introduce some corrections to Equation (54) [135], leading to the effective bending modulus, which is smaller than the intrinsic bending modulus due to the presence of the additional degree of freedom (composition–curvature coupling). The comparison of the bending moduli obtained from the analysis of shape fluctuations with that obtained employing the stress–strain relation of lipid tubes, Equation (31), shows that the former is always smaller than the latter [89]. In Ref. [89], it was also shown that it is possible to determine the composition-curvature coupling in fluctuating binary lipid membranes: $\Delta \varphi = \Lambda K$, where $\Delta \varphi$ is a local composition difference between the opposing monolayers, and $K$ is the curvature of the midsurface. It is also possible to obtain a direct analytical expression for $\Lambda$, which is given by a rather bulky Equation (14) of Ref. [89] and which depends, among other things, on the spontaneous curvature of monolayers, making possible its determination from measuring $\Lambda$.

The shape fluctuations can be analyzed not only for the planar configuration. Another rather simple configuration is a lipid nanotube. The shape $u$ of a cylindrical tube can be parametrized by cylindrical coordinates, $0 \leq \varphi \leq 2\pi$ and $0 \leq \zeta \leq L/R$, where $L$ and $R$ are the tube length and radius, respectively:

$$\mathbf{r}(\varphi,\zeta) = \begin{pmatrix} x \\ y \\ z \end{pmatrix} = R \begin{pmatrix} [1+u(\phi,\zeta)]\cos\varphi \\ [1+u(\phi,\zeta)]\sin\varphi \\ \zeta \end{pmatrix}. \tag{56}$$

The Fourier amplitudes of the Fourier series $u(\varphi,\zeta) = \sqrt{\dfrac{R}{2\pi L}} \sum_{m,\bar{q}} u_{m,\bar{q}} e^{i(m\varphi+\bar{q}\zeta)}$ satisfy the following relation [136]:

$$\left\langle \left|u_{m,\bar{q}}\right|^2 \right\rangle = \frac{k_B T}{kQ^4}, \tag{57a}$$

$$Q^4 = (m^2 - 1)^2 + \bar{q}^2(\bar{q}^2 + 2m^2), \tag{57b}$$

where $m \in \mathbb{Z}$ and $\bar{q} = 2\pi n R / L$, $n \in \mathbb{Z}$. In Ref. [88], the analysis of the shape fluctuations of 2D meshless membrane tubes showed that at low $Q$ values, Equation (57a) well describes the fluctuations, and the bending modulus value determined based on Equation (57a) agrees with that of planar membranes obtained based on Equation (54). At high $Q$ values, however, a softening similar to that of planar membranes also occurs, which can be attributed to the protrusion mode.

*4.4. Fluctuations of Director*

Due to the thermal motion, not only the membrane shape fluctuates but also the orientation of individual lipid molecules. In Section 2, it was shown that within the 3D classical theory of elasticity, the orientations of lipid molecules are characterized by the vector field of unit vectors $\mathbf{n}$, called directors. The elastic energy written in terms of $\mathbf{n}$ is given either by the HK Hamiltonian, Equation (4), or by a more general one, Equation (6), where $\mathbf{n}$ enters through the effective curvature as $\tilde{K} \approx \nabla \cdot \mathbf{n}$, where $\nabla \cdot \mathbf{n}$ is the divergence of $\mathbf{n}$. Within the framework of the HK Hamiltonian, it can be shown that the director fluctuations of planar lipid bilayers satisfy the following relation [137]:

$$\left\langle \left|\hat{n}_{\mathbf{q}}^{\parallel}\right|^2 \right\rangle = \frac{k_B T}{K_c q^2}, \tag{58}$$

where $K_c$ is the bilayer bending modulus and $\hat{n}_{\mathbf{q}}^{\parallel} = \dfrac{\hat{\mathbf{n}}_{\mathbf{q}} \cdot \mathbf{q}}{q}$, where $\hat{\mathbf{n}}_{\mathbf{q}}$ is the Fourier amplitude of $\hat{\mathbf{n}} = \dfrac{1}{2}\left[\mathbf{n}^u - \mathbf{n}^l\right]$ with $\mathbf{n}^u$ and $\mathbf{n}^l$ being the lipid director fields of the upper and lower monolayers, respectively. In Refs. [137,138], it was shown that $\left\langle \left| \hat{n}_{\mathbf{q}}^{\|} \right|^2 \right\rangle$ appears to be a more useful quantity than $\left\langle \left| h_{\mathbf{q}} \right|^2 \right\rangle$ for the determination of the bending modulus, as $q^2 \left\langle \left| \hat{n}_{\mathbf{q}}^{\|} \right|^2 \right\rangle$ exhibits a plateau regime at higher values of $q$ than $q^4 \left\langle \left| h_{\mathbf{q}} \right|^2 \right\rangle$, which allows for a more accurate determination of the bending modulus for small systems, reducing the simulation time. Still, at high $q$ values, there is a substantial discrepancy between theory and simulation data: $q^2 \left\langle \left| \hat{n}_{\mathbf{q}}^{\|} \right|^2 \right\rangle$ appears to be a monotonically increasing function of $q$ [137], not a constant as Equation (58) predicts. In Refs. [40,41], it was proposed that the tilt–curvature coupling term $\sim \mathbf{T} \cdot \nabla \tilde{K}$ is responsible for this discrepancy: the inclusion of this term into the elastic energy modifies the fluctuation spectrum to a monotonically increasing function. However, in Ref. [14], it was shown that the curvature gradient term, $\sim \left( \nabla \tilde{K} \right)^2$, should also be included in the elastic energy to provide the overall stability of a lipid membrane. With the curvature gradient included in the elastic energy, the theoretical prediction of the fluctuation spectrum for $q^2 \left\langle \left| \hat{n}_{\mathbf{q}}^{\|} \right|^2 \right\rangle$ transforms to a monotonically decreasing function of $q$. Therefore, still the discrepancy between the theory and fluctuation data from MD simulations exist, which might indicate the failure of the elastic theory at high $q$ values and a necessity to include protrusion-like deformation modes to explain the fluctuation spectrum. In this respect, the determination of the elastic moduli of small-scale deformation modes, such as the tilt modulus or tilt–curvature coupling modulus, which require the sampling at high $q$ values, appears to be problematic. Another difficulty with director fluctuations is ambiguity in the definition of a lipid director. As with shape fluctuations, a set of reference beads should be chosen to define the lipid director field. It was shown that the ambiguity in the choice of these reference beads can lead to a rather large systematic error up to 40% (according to row 2 of Figure 5 in Ref. [134]).

Fluctuations of lipid directors can be considered not only in the Fourier space. In Refs. [139–141], a method (termed RSF for real-space fluctuations) was proposed wherein the director fluctuations are analyzed in real space. It was shown that the monolayer bending modulus can be determined from the probability distribution of an angle $\alpha$ between the directors of neighboring lipids [140]:

$$-\frac{2k_B T}{a_0} \ln \left[ \frac{P(\alpha)}{\sin \alpha} \right] = k_m \alpha^2 + C, \tag{59}$$

where $C$ is a constant and $a_0$ is the area per lipid. Equation (59) is valid only for small tilt angles, $\leq 10°$ [140]. A quadratic fit to Equation (59) thus yields the monolayer bending modulus $k_m$. In addition to the bending modulus, the tilt modulus can be determined analogously by employing the following equation [141]:

$$-\frac{4k_B T}{a_0} \ln \left[ \frac{P(\theta)}{\sin \theta} \right] = k_t \theta^2 + C, \tag{60}$$

where $\theta$ is the angle between the lipid director and local monolayer normal.

RSF can be applied not only to single-component but also to multicomponent monolayers: $P(\alpha)$ and $P(\theta)$ can be calculated for every pair of lipid names and then an averaging performed [140]. Although a particular averaging type is a matter of choice and is not strictly justified, it was shown that the bending moduli obtained by the Reuss averaging [142] agree well with experimental data [140]. It is of note, however, that uncertainties in the experimental data, with which the results of Ref. [140] were compared, are quite large. In Ref. [143], using RSF among other things, it was shown that cholesterol stiffens lipid membranes composed not only of saturated lipids but also of unsaturated ones, which shed light on the understanding of cholesterol's role in membrane functions, such as viral budding. The Reuss averaging employed in RSF has the following form:

$$\frac{1}{k_m} = \frac{1}{\varphi_{tot}} \sum_{i,j} \frac{\varphi_{ij}}{k_m^{ij}}, \quad (61)$$

where $i$ and $j$ indicate a particular lipid type, $k_m^{ij}$ is the bending modulus of the $ij$th pair, $\varphi_{ij}$ is the average number of near neighbor pairs between lipids of type $i$ and $j$, and $\varphi_{tot}$ is the total number of pairs. Equation (61), thus, involves the averaging of the inverse bending moduli. An alternative to the Reuss averaging exists, the Voigt averaging [142], according to which the bending moduli are averaged directly, not their inverse values. Both the Voigt averaging and Reuss averaging apply to the Young's modulus, which quantifies the energetic cost of uniform stretching of composite materials [142], while the applicability of these averaging approaches to the bending modulus of lipid mixtures is uncertain. Sometimes, the Reuss averaging is applied in some other form [144,145]. For example, in Ref. [144], the following form of the Reuss averaging was employed:

$$\frac{1}{k_m} = \left( \sum_i a_i N_i \right)^{-1} \left( \sum_i \frac{a_i N_i}{k_{m,i}} \right), \quad (62)$$

where $N_i$ is the number of lipids of type $i$, $a_i$ and $k_{m,i}$ are the area per lipid and bending modulus of single-component monolayers composed of lipids of type $i$, respectively. However, in Ref. [43], it was shown that Equation (62) does not hold for a binary mixture of DPPC and DOPC, which indicates that considerations based on the Reuss averaging may not always apply to lipid mixtures.

*4.5. Virtual Deformations*

When lipid membranes fluctuate, many deformation modes are simultaneously involved. However, by a straightforward application of the rules of statistical mechanics, it is possible to single out a deformation mode of interest and study the statistics of this deformation mode, thereby virtually imposing this deformation on a membrane [84,129]. In particular, each deformation mode can be characterized by some perturbation variable $\lambda$, which characterizes the corresponding strain of a deformation. $\lambda$ may be, for instance, lateral stretching, curvature, tilt angle, etc. The free energy $F_\lambda$ corresponding to strain $\lambda$ can be expressed as $F_\lambda = \frac{1}{2} k_\lambda \lambda^2$, where $k_\lambda$ is the elastic modulus corresponding to $\lambda$. Thus, the corresponding elastic stress, $S_\lambda$, and $k_\lambda$ can be expressed as $S_\lambda = \frac{\partial F_\lambda}{\partial \lambda}$ and $k_\lambda = \frac{\partial^2 F_\lambda}{\partial \lambda^2}$, respectively. On the other hand, $S_\lambda$ and $k_\lambda$ can be expressed through the overall free energy $F$ of a system [84]:

$$F = -k_B T \ln \left( \int e^{-E/k_B T} dE \right), \quad (63a)$$

$$S_\lambda = \left. \frac{\partial F}{\partial \lambda} \right|_{V,T} = \frac{\int \frac{\partial E}{\partial \lambda} e^{-E/k_B T} dE}{\int e^{-E/k_B T} dE} = \left\langle \frac{\partial E}{\partial \lambda} \right\rangle_{V,T}, \quad (63b)$$

$$k_\lambda = \left. \frac{\partial^2 F}{\partial \lambda^2} \right|_{V,T} = \left\langle \frac{\partial^2 E}{\partial \lambda^2} \right\rangle_{V,T} - \frac{1}{k_B T} \left( \left\langle \left( \frac{\partial E}{\partial \lambda} \right)^2 \right\rangle_{V,T} - \left\langle \frac{\partial E}{\partial \lambda} \right\rangle_{V,T}^2 \right), \quad (63c)$$

where $E$ is the energy of a system. As the kinetic energy provides only a constant to $F$, $E$ in Equation (63b,c) can be replaced with the potential energy $U(\mathbf{r}_1,...,\mathbf{r}_N)$, where $\mathbf{r}_i$ are the position vectors of all $N$ particles of a system. Let us consider, for example, the lateral stretching deformation mode with stretching $\alpha$. The energy is given by $A\dfrac{k_A \alpha^2}{2}$ and $\lambda = \alpha$. The stretching deformation can be parametrized as:

$$\mathbf{r}'_i = \begin{pmatrix} 1+\dfrac{\varepsilon}{2} & 0 & 0 \\ 0 & 1+\dfrac{\varepsilon}{2} & 0 \\ 0 & 0 & 1-\varepsilon \end{pmatrix} \mathbf{r}_i . \tag{64}$$

Employing Equations (63b) and (64), we can write that the lateral tension, $\sigma = k_A \alpha$, can be expressed as:

$$\sigma = \frac{1}{A} \sum_i \left\langle \frac{1}{2}\left( x_i \frac{\partial U}{\partial x_i} + y_i \frac{\partial U}{\partial y_i} \right) - z_i \frac{\partial U}{\partial z_i} \right\rangle , \tag{65}$$

where $x_i$, $y_i$ and $z_i$ are the coordinates of the position vector $\mathbf{r}_i$.

In Ref. [84], a method was proposed for the calculation of the bending modulus and Gaussian curvature modulus employing Equation (63c). For this, two virtual deformations were considered: spherical and cylindrical with the corresponding mean curvature radii of $1/C_{sp}$ and $1/C_{cy}$. As, according to the Helfrich Hamiltonian, Equation (1), the elastic energy of these deformations equals $\dfrac{k}{2}(C_{cy} - C_0)^2 A$ and $\left[\dfrac{k}{2}(2C_{sp} - C_0)^2 + \bar{k} C_{sp}^2\right] A$, the elastic moduli can be expressed as [84]:

$$k = \frac{1}{A} \frac{\partial^2 F}{\partial C_{cy}^2} \bigg|_{V,T,C_{cy}=0} , \tag{66a}$$

$$4k + 2\bar{k} = \frac{1}{A} \frac{\partial^2 F}{\partial C_{sp}^2} \bigg|_{V,T,C_{sp}=0} . \tag{66b}$$

The corresponding expressions through the potential energy $U$, which can be obtained after substituting the parametrizations of the spherical and cylindrical deformations into Equation (63b,c), are too bulky to be presented here and are given by Equations (26) and (27) of Ref. [84]. It was shown that the method works well for 1D meshless membranes [84]. However, for thick membranes, a further extension of the theory is required to take into account the volume fluctuations effects, as these volume fluctuations enter the parameterizations of the spherical and cylindrical deformations [84]. It was also pointed out that for the calculation of the derivatives of Equation (63c), a choice of the geometrical center is required, the definition of which contains some ambiguity. Nevertheless, the advantage of the approach of Refs. [84,129] is that it does not require performing the force decomposition, which also leads to some ambiguity in the local-stress methods [84].

## 5. Discussion

The development of MD led to the appearance of *in silico* approaches for the determination of elastic parameters of simulated lipid membranes. In this work, a review of these approaches, with a focus on theoretical aspects, was provided. All approaches were divided into two broad groups: equilibrium force methods and fluctuation-based methods. In the first group of methods, average equilibrium quantities of stress and strain are measured, while in the second group, the deviations from the average values are analyzed.

In MD simulations, it is possible to measure the positions and velocities of individual atoms, which permits a more detailed consideration of the lipid membrane mechanics unattainable in experiments. At the same time, this detailed characteristic of MD simulations also poses a difficulty. In particular, a transition from the continuum language of elastic theories to the discrete representation of MD simulations is required. A way to accomplish this transition is not straightforward. A major challenge exists regarding the description of discrete quantities, which often leads to ambiguities. In equilibrium force approaches, for instance, the definition of the local stress is not unique (see Section 3.2.2) [83]. As for the fluctuation-based methods, the problems are with the definitions of lipid directors and membrane shapes (see Sections 4.3–4.4), which leads to significant systematic errors [134]. Therefore, one has to rely on reasonable choices [82,134]. In addition, at small deformation scales comparable with membrane thickness, the fluctuation-based methods show the discrepancy between theory and simulation data [14], which might indicate the failure of the elastic theory at short fluctuation wavelengths and which makes the determination of the elastic moduli of small-scale deformation modes, such as the tilt modulus or the tilt–curvature coupling modulus, problematic. These issues represent an important problem to be solved in subsequent developments of MD methods.

Another challenge concerns the determination of elastic parameters of lipid mixtures. Elastic parameters of lipid mixtures are of particular interest since cell membranes are multicomponent [26,27]. However, deformations of such membranes involve the composition–curvature coupling effect. This implies that curved geometries with nonconstant extrinsic curvature, such as in the buckling procedure [96,98], are not applicable to the determination of intrinsic bending moduli and spontaneous curvatures. The same is true for the cylindrical geometry of tubular bilayers, in which the curvature radii of outer and inner monolayers are different [91]. In addition, the composition–curvature coupling is present in fluctuation-based methods [135], which also impedes the determination of intrinsic elastic parameters of multicomponent lipid membranes. The constant curvature geometry of planar bilayers still enables the determination of intrinsic parameters. However, for their determination, the calculation of the local stress is required, the definition of which is still in the state of no consensus. It is also possible to look at the local real-space fluctuations of lipid directors [140] to determine the intrinsic bending moduli of lipid mixtures. The latter approach, however, requires the choice of averaging made for pairwise splay moduli, which is not strictly justified. Moreover, it was shown that a particular type of averaging employed in Ref. [140], the Reuss averaging, may not always hold for lipid mixtures.

In conclusion, although a significant development of MD approaches for the determination of elastic parameters of lipid membranes has occurred since the advent of the MD, several problems still exist with the application of these approaches. In particular, two main challenges have been identified: (i) the ambiguity in the transition from the continuum description of elastic theories to the discrete representation of MD simulations, and (ii) the determination of intrinsic elastic parameters of lipid mixtures, which is complicated due to the composition–curvature coupling effect. Further development of the MD approaches for the determination of lipid membrane elastic parameters will contribute to obtaining new insights into the biological role of mechanical properties of lipid membranes and their lipid composition in the life cycle of living cells.

**Funding:** This research was funded by the Ministry of Science and Higher Education of the Russian Federation (grant agreement # 075-15-2020-782).

**Institutional Review Board Statement:** Not applicable.

**Data Availability Statement:** Not applicable.

**Acknowledgments:** Grateful acknowledgment to Dr. Timur R. Galimzyanov for providing critical comments on the manuscript.

**Conflicts of Interest:** The author declares no conflict of interest.